\documentclass[onecolumn]{article}

\usepackage[a4paper, total={6.5in, 8.5in}]{geometry}
\usepackage[utf8]{inputenc}

\usepackage[square,numbers]{natbib}
\usepackage{amsmath,amsfonts}
\usepackage{mathtools}
\usepackage{tikz}
\usepackage{graphicx}
\usepackage{subcaption}

\usetikzlibrary{arrows}
\usepackage[inline]{enumitem}
\usepackage{comment}
\usepackage{dsfont}

\usepackage{algorithm2e}
\makeatletter
\renewcommand{\@algocf@capt@plain}{above}
\makeatother
\makeatletter
\newcommand{\removelatexerror}{\let\@latex@error\@gobble}
\makeatother

\usepackage[labelformat=simple]{subcaption}

\newcommand{\BlackBox}{\rule{1.5ex}{1.5ex}}  % end of proof
\ifdefined\proof
    \renewenvironment{proof}{\par\noindent{\bf Proof\ }}{\hfill\BlackBox\\[2mm]}
\else
    \newenvironment{proof}{\par\noindent{\bf Proof\ }}{\hfill\BlackBox\\[2mm]}
\fi

\newcommand{\bulletarrow}{
  \setlength{\unitlength}{1mm}
  \begin{picture}(5,1)(0,0)
    \put(0.2,0){$\bullet$}
    \put(1,0){$\rightarrow$}
  \end{picture}
}

\newcommand{\arrowbullet}{
  \setlength{\unitlength}{1mm}
  \begin{picture}(5,1)(0,0)
    \put(0.2,0){$\leftarrow$}
    \put(3,0){$\bullet$}
  \end{picture}
}

\newcommand{\circarrow}{
  \setlength{\unitlength}{1mm}
  \begin{picture}(5,1)(0,0)
    \put(1,1){\circle{1}}
    \put(1.2,0){$\rightarrow$}
  \end{picture}
}

\newcommand{\circcirc}{
  \setlength{\unitlength}{1mm}
  \begin{picture}(5,1)(0,0)
    \put(1,1){\circle{1}}
    \put(1.5,1){\line(1,0){2}}
    \put(4,1){\circle{1}}
  \end{picture}
}

\newtheorem{example}{Example} 
\newtheorem{theorem}{Theorem}
\newtheorem{lemma}[theorem]{Lemma} 
\newtheorem{definition}[theorem]{Definition}

\DeclareMathOperator{\An}{An}
\DeclareMathOperator{\Pa}{Pa}
\DeclareMathOperator{\pa}{pa}
\DeclareMathOperator{\De}{De}
\DeclareMathOperator{\Adj}{Adj}

\newcommand{\mb}[1]{\mathbf{#1}}
\newcommand{\dsepp}{\perp_{d}}
\newcommand{\msepp}{\perp_{m}}
\newcommand{\g}[1][G]{\mathcal{#1}}
\newcommand*\diff{\mathop{}\!\mathrm{d}}
\newcommand{\ind}{\perp\!\!\!\!\perp}
\newcommand{\notind}{\not\!\perp\!\!\!\!\perp}

\newenvironment{proofof}[1][]{\noindent \textbf{Proof of #1.}}{\hfill\BlackBox}
\newenvironment{proofofnoqed}[1][]{\noindent \textbf{Proof of #1.}}{}
\newenvironment{proofsketch}{\noindent \textbf{Proof Sketch.}}{\hfill\BlackBox\\}

\usepackage{color}
\definecolor{blue}{rgb}{0.0, 0.5, 0.5}

\title{Data-Driven Adjustment for Multiple Treatments}
\author{Sara LaPlante (slap47@uw.edu) \\
        \small Department of Statistics, University of Washington, USA \\ \\ 
        Sofia Triantafillou (sof.triantafillou@uoc.gr) \\
        \small Department of Mathematics \& Applied Mathematics, University of Crete, Greece\\
        \small Institute of Applied and Computational Mathematics, FORTH - Hellas, Greece\\ \\
        Emilija Perkovi\'c (perkovic@uw.edu) \\
        \small Department of Statistics, University of Washington, USA \\ \\
}
\date{}

\begin{document}
\maketitle

% ======================================================
%
%                        MAIN TEXT
%
% ======================================================

%%%%%%%%%%%%%%%%%%%%%%%%%%%%%%%%%%%%
%              ABSTRACT            %
%%%%%%%%%%%%%%%%%%%%%%%%%%%%%%%%%%%%

\begin{abstract}
{Covariate adjustment is one method of causal effect identification in non-experimental settings. Prior research provides routes for finding appropriate adjustments sets, but much of this research assumes knowledge of the underlying causal graph. In this paper, we present two routes for finding adjustment sets that do not require knowledge of a graph -- and instead rely on dependencies and independencies in the data directly. We consider a setting where the adjustment set is unaffected by treatment or outcome. The first route shows how to extend prior research in this area using a concept known as \textit{c-equivalence}. Our second route provides sufficient criteria for finding adjustment sets in the setting of multiple treatments.}
\end{abstract}

%%%%%%%%%%%%%%%%%%%%%%%%%%%%%%%%%%%%
%              INTRO               %
%%%%%%%%%%%%%%%%%%%%%%%%%%%%%%%%%%%%

\section{Introduction}
\label{sec:intro}

To identify a causal effect from observational data, researchers often turn to covariate adjustment, which can eliminate concerns of confounding bias. But choosing a set of adjustment variables that will accurately identify the causal effect of interest requires carefulness. Much of the literature has sought routes to finding such a set, and these routes typically include two steps: (1) knowing or learning a causal graph and (2) checking sets of variables in the graph against a list of graphical criteria.

For example, Pearl \cite{pearl1995causal} introduced graphical requirements known as the \textit{back-door criterion} for use when a causal \textit{directed acyclic graph} (DAG) is known. This criterion is sufficient for identifying the effect of a single treatment through adjustment. Subsequently, a graphical \textit{adjustment criterion} for DAGs was developed that is both necessary and sufficient for identifying the effect of multiple treatments \cite{shpitser2012validity, perkovic2015complete,perkovic2018complete}. Further results have considered settings where a full DAG is not known, including necessary and sufficient graphical criteria
for \textit{maximal ancestral graphs} (MAGs), which allow for latent confounding \citep{van2014constructing}, and extensions to equivalence classes of DAGs and MAGs known as \textit{completed partially directed acyclic graphs} (CPDAGs) and \textit{partial ancestral graphs} (PAGs) \citep{maathuis2015generalized, perkovic2017interpreting, perkovic2018complete}, respectively. Notably, one of these extensions -- the \textit{generalized adjustment criterion} \cite{perkovic2018complete} -- is necessary and sufficient for identification in all such graphs.

In their \citeyear{entner2013data} paper, Entner, Hoyer, and Spirtes (EHS) \cite{entner2013data} also consider identifying causal effects through covariate adjustment. But unlike research that relies on graphical criteria, EHS focus on identification through the observed data directly. Their paper's main result is a pair of rules that they show are necessary and sufficient for discovering when a causal effect is identifiable. The first of these rules -- reproduced as Theorem \ref{thm:r1-entner} below -- provides an adjustment set for identifying a causal effect when one exists. The strength of this data-driven rule lies in its simplicity: the researcher only needs to find an observed variable that matches two conditional in/dependence criteria.

We consider extending the results of EHS \citep{entner2013data} -- first by reviewing the notion of c-equivalence \cite{pearl2014confounding}. Notably, any set that is c-equivalent to an adjustment set must also be an adjustment set. So while the rules of EHS \cite{entner2013data} guarantee finding one adjustment set when the causal effect is identifiable, subsequently finding c-equivalent sets will uncover additional sets for adjustment. Pearl and Paz \citep{pearl2014confounding} provide criteria sufficient for finding c-equivalent sets. And since these criteria are based on in/dependencies in the data directly, they can be used to extend the EHS \cite{entner2013data} criterion, without requiring additional graphical assumptions. We note briefly (for further discussion below) that having more than one adjustment set may seem unnecessary for practical research. However, this choice can be crucial in the process of causal effect estimation.

Our main contribution is an extension of EHS \cite{entner2013data} to a setting with multiple treatments. That is, where EHS \cite{entner2013data} consider only a single treatment $X$, we consider a set of treatments $\mb{X}$. We develop two data-driven rules, analogous to the first rule of EHS \cite{entner2013data} (Theorem \ref{thm:r1-entner}), that are sufficient for finding adjustment sets in this setting. Our first rule (Theorem \ref{thm:r1-build}) finds an adjustment set for $\mb{X}$ by finding a causal ordering of the treatments and building up from an adjustment set for the first treatment in the causal ordering. Our second rule (Theorem \ref{thm:r1-combine}) finds an adjustment set for $\mb{X}$ by combining adjustment sets for each $X \in \mb{X}$ after paring off unnecessary variables. This process relies on the notion of c-equivalence.

This paper is organized as follows. Section \ref{sec:prelims} provides a set of definitions for graphical models. Section \ref{sec:extension1} explains how c-equivalence can be used to extend the results of EHS \cite{entner2013data}. Section \ref{sec:r1} details our extension of EHS \cite{entner2013data} to multiple treatments. Then we measure the performance of our rules through a data simulation in Section \ref{sec:simulations}, and we discuss the limitations of our work and suggestions for future research in Section \ref{sec:discussion}.

%%%%%%%%%%%%%%%%%%%%%%%%%%%%%%%%%%%%
%             PRELIMS              %
%%%%%%%%%%%%%%%%%%%%%%%%%%%%%%%%%%%% 

\section{Preliminaries}
\label{sec:prelims}

%----------------  DAGS  ----------------%
Throughout this paper, we assume a causal model that induces a directed graph. The following are key definitions related to these graphs and their associated densities. We rely on the framework of Pearl \cite{pearl2009causality}.

\textbf{Nodes, Edges, and Graphs.}
We use capital letters (e.g., $X$) to denote nodes in a graph as well as the random variables these nodes represent. We use bold capital letters (e.g., $\mb{X}$) to denote node sets. A \textit{graph} $\g=(\mb{V},\mb{E})$ consists of a set of nodes $\mb{V}$ and a set of edges $\mb{E}$. A \textit{directed graph} contains only directed edges ($\to$).

\textbf{Paths and DAGs.}
For disjoint node sets $\mb{X}$ and $\mb{Y}$, a \textit{path} from $\mb{X}$ to $\mb{Y}$ is a sequence of distinct nodes $\langle X, \dots,Y \rangle$ from some $X \in \mb{X}$ to some $Y \in \mb{Y}$ for which every pair of successive nodes is adjacent. A \textit{directed path} from $X$ to $Y$ is a path of the form $X \to \dots \to Y$. A directed path from $X$ to $Y$ and the edge $Y \to X$ form a \textit{directed cycle}. A directed graph without directed cycles is a \textit{directed acyclic graph} (DAG). 

\textbf{Colliders and Non-colliders.}
The \textit{endpoints} of a path $p = \langle X_1, \dots, X_k \rangle$ are the nodes $X_1$ and $X_k$. For $1 < i < k$, the node $X_i$ is a \textit{collider} on $p$ if $p$ contains $X_{i-1} \to X_i \gets X_{i+1}$, and $X_i$ is a \textit{non-collider} on $p$ if $p$ contains $X_{i-1} \gets X_i$ or $X_i \to X_{i+1}$.

\textbf{Ancestral Relationships.}
If $X \to Y$, then $X$ is a \textit{parent} of $Y$. If there is a directed path from $X$ to $Y$, then $X$ is an \textit{ancestor} of $Y$ and $Y$ is a \textit{descendant} of $X$. We use the convention that every node is an ancestor and descendant of itself. The sets of parents, ancestors, and descendants of $X$ in $\g[D]$ are denoted by $\Pa(X,\g[D])$, $\An(X,\g[D])$, and $\De(X,\g[D])$, respectively. We let $\An(\mb{X},\g[D]) = \cup_{X \in \mb{X}} \An(X,\g[D])$ and $\De(\mb{X},\g[D]) = \cup_{X \in \mb{X}} \De(X,\g[D])$. Unconventionally, we define $\Pa(\mb{X},\g[D]) =[\, \cup_{X \in \mb{X}} \Pa(X,\g[D]) \,] \setminus \mb{X}$. 

\textbf{Markov Compatibility and Faithfulness.}
An \textit{observational density} $f(\mb{v})$ is \textit{Markov compatible} with a DAG $\g[D] = (\mb{V},\mb{E})$ if $f(\mb{v})= \prod_{V_i \in \mb{V}}f(v_i|\pa(v_i,\g[D]))$. It is \textit{faithful} to $\g[D]$ if ${(\mb{X} \ind \mb{Y} \,|\, \mb{Z})_f}$ implies ${(\mb{X} \dsepp \mb{Y} \,|\, \mb{Z})_{\g[D]}}$ (see definition of \textit{d-separation} below). We require $f(\mb{v})>0$ for all valid values of $\mb{V}$.

\textbf{D-connection, D-separation, and Probabilistic Implications.}
Let $\mb{X}$, $\mb{Y}$, and $\mb{Z}$ be pairwise disjoint node sets in a DAG $\g[D]$. A path $p$ from $\mb{X}$ to $\mb{Y}$ is \textit{d-connecting} (or \textit{open}) given $\mb{Z}$ if every non-collider on $p$ is not in $\mb{Z}$ and every collider on $p$ has a descendant in $\mb{Z}$. Otherwise, $p$ is \textit{blocked} given $\mb{Z}$. If all paths between $\mb{X}$ and $\mb{Y}$ in $\g[D]$ are blocked given $\mb{Z}$, then $\mb{X}$ is \textit{d-separated} from $\mb{Y}$ given $\mb{Z}$ in $\g[D]$ and we write $(\mb{X} \dsepp \mb{Y} | \mb{Z})_{\g[D]}$. This d-separation implies that $\mb{X}$ and $\mb{Y}$ are independent given $\mb{Z}$ in any \textit{observational density} that is Markov compatible with $\g[D]$ \citep{lauritzen1990independence}. 

%----------------

\textbf{Causal Graphs.}
Let $\g[D]$ be a DAG with nodes $V_i$ and $V_j$. Then $\g[D]$ is a \textit{causal DAG} if every edge $V_i \to V_j$ represents a direct causal effect of $V_i$ on $V_j$. In a causal DAG, any directed path is \textit{causal}, and any other path is \textit{non-causal}.

\textbf{Consistency.}
Let $f(\mb{v})$ be an observational density over $\mb{V}$. The notation $do(\mb{X} = \mb{x})$, or $do(\mb{x})$ for short, represents an outside intervention that sets $\mb{X} \subseteq \mb{V}$ to fixed values $\mb{x}$. An \textit{interventional density} $f(\mb{v}|do(\mb{x}))$ is a density resulting from such an intervention.

Let $\mb{F^*}$ denote the set of all interventional densities $f(\mb{v}|do(\mb{x}))$ such that $\mb{X} \subseteq \mb{V}$ (including $\mb{X} = \emptyset$). A causal DAG $\g[D] = (\mb{V,E})$ is a \textit{causal Bayesian network compatible with} $\mb{F^*}$ if and only if for all $f(\mb{v} | do(\mb{x})) \in \mb{F^*}$, the following \textit{truncated factorization} holds:
\begin{align*}
    f(\mb{v} | do(\mb{x})) = \prod_{V_i \in \mb{V} \setminus \mb{X}} f(v_i|\pa(v_i,\g[D])) \mathds{1}(\mb{X} = \mb{x}).
\end{align*}
We say an interventional density is \textit{consistent} with a causal DAG $\g[D]$ if it belongs to a set of interventional densities $\mb{F^*}$ such that $\g[D]$ is compatible with $\mb{F^*}$. Note that any observational density that is Markov compatible with $\g[D]$ is consistent with $\g[D]$.

\textbf{Causal Models.}
A \textit{structural equation model} (SEM), or \textit{causal model}, is a set of equations -- one for each random variable $V \in \mb{V}$ that maps the causal determinants of $V$, along with random noise, to the values of $V$. This model induces a DAG $\g[D]$ over $\mb{V}$ and a set of interventional densities $\mb{F^*} = \{f(\mb{v}|do(\mb{x})): \mb{X} \subseteq \mb{V}\}$ that are consistent with $\g[D]$. The joint density $f(\mb{v}) \in \mb{F^*}$ is faithful to $\g[D]$. 

\textbf{Identifiability.}
Let $\mb{X}$ and $\mb{Y}$ be disjoint node sets in a causal DAG $\g[D] = (\mb{V,E})$ that is compatible with $\mb{F^*_i} = \{ f_i(\mb{v}|do(\mb{x'})) : \mb{X'} \subseteq \mb{V} \}$. We say the causal effect of $\mb{X}$ on $\mb{Y}$ is \textit{identifiable} in $\g[D]$ if for any $\mb{F^*_1}, \mb{F^*_2}$ where $f_1(\mb{v}) = f_2(\mb{v})$, we have $f_1(\mb{y}|do(\mb{x})) = f_2(\mb{y}|do(\mb{x}))$.

\textbf{Adjustment Sets.}
Let $\mb{X}$, $\mb{Y}$, and $\mb{Z}$ be pairwise disjoint node sets in a causal DAG $\g[D]$. Then $\mb{Z}$ is an adjustment set relative to $(\mb{X,Y})$ in $\g[D]$ if and only if $f(\mb{y}|do(\mb{x})) = \int f(\mb{y} | \mb{x}, \mb{z}) f(\mb{z}) \diff \mb{z}$ for any $f$ consistent with $\g[D]$. We omit reference to $(\mb{X,Y})$ or $\g[D]$ when it can be assumed.

\textbf{Causal Ordering.}
Let $\mb{V}=\{V_1, \dots, V_k\}$, $k \ge 1$, be a set of random variables in a causal model. We say ${V_1 < \dots < V_k}$ is a \textit{causal ordering consistent with the model} if $V_j$ is not a causal ancestor of $V_i$ for $1 \le i < j \le k$. Note there can be more than one causal ordering. For example, a causal model that induces the DAG $X_i \leftarrow X_j \rightarrow X_k$, has consistent orderings $X_j < X_i < X_k$ and $X_j < X_k < X_i$ and $X_j < \{X_i, X_k\}$.

%----------------  PAGS  ----------------%

\textbf{PAGs.}
We reference \textit{partial ancestral graphs} (PAGs; \citep{richardson2002ancestral}) in several examples of Sections \ref{sec:extension1}-\ref{sec:r1} and the simulations of Section \ref{sec:simulations}. However, our results require no knowledge of PAGs directly. Thus, we suppress related definitions to Supp. \ref{supp:prelims} and provide an informal overview below.

We can represent a causal model that has unmeasured variables by using a \textit{maximal ancestral graph} (MAG) over the observed variables alone. Directed edges ($\to$) in a MAG denote causal ancestry, and bi-directed edges ($\leftrightarrow$) denote the presence of an unmeasured confounder. MAGs encode all the conditional in/dependencies among observed variables through a graphical criterion called \textit{m-separation}.

PAGs represent an equivalence class of MAGs with the same m-separations. Directed and bi-directed edges in a PAG denote shared ancestry and confounding, respectively, among all represented MAGs. Circle edge marks denote disagreement among represented MAGs. For example, $X \circarrow Y$ denotes that at least one represented MAG has the edge $X \to Y$ and at least one represented MAG has $X \leftrightarrow Y$.

%%%%%%%%%%%%%%%%%%%%%%%%%%%%%%%%%%%%
%      RESULTS - C-EQUIVALENCE     %
%%%%%%%%%%%%%%%%%%%%%%%%%%%%%%%%%%%%

\section{Insights from Existing Work}
\label{sec:extension1}

In this section, we review the first rule of EHS \cite{entner2013data} and suggest a known equivalency of adjustment sets as a direct extension. We close by providing a rationale for why this extension would be useful for causal effect estimation.

%------------------------------------
%  REVIEWING R1
%------------------------------------

\subsection{The Original Rule}

As noted in Section \ref{sec:intro}, EHS \cite{entner2013data} develop the following rule for finding an adjustment set when the causal effect of a single treatment is identifiable. 

\begin{theorem}
\label{thm:r1-entner}
{\normalfont (\textbf{R1 Entner})}
    Let $\{X\}$, $\{Y\}$, and $\mb{W}$ be pairwise disjoint sets of observed random variables in a causal model. Suppose ${\mb{W} < X < Y}$ is a causal ordering consistent with the model. If there exists $W \in \mb{W}$ and $\mb{Z} \subseteq \mb{W} \setminus \{W\}$ such that 
    \begin{enumerate}[label=(\roman*)]
    	\item $W \notind Y \,\,|\,\, \mb{Z}$ \,\, and \label{thm:r1-entner-i}
    	\item $W \ind Y \,\,|\,\, \mb{Z} \cup \{X\}$, \label{thm:r1-entner-ii}
    \end{enumerate}
    then $X$ has a causal effect on $Y$ that is identifiable through the adjustment set $\mb{Z}$.
\end{theorem}

EHS \cite{entner2013data} show that the rule above is necessary for identification (see their Theorem 3). That is, if the causal effect of $X$ on $Y$ is identifiable and nonzero, then Theorem \ref{thm:r1-entner} will find an adjustment set. However, we note that Theorem \ref{thm:r1-entner} only guarantees finding one such set. In some cases, such as in Example \ref{ex:no-entner1} below, there may be additional adjustment sets that cannot be found using Theorem \ref{thm:r1-entner}.

%----------------------------
% Example - No Entner 1
%----------------------------

\begin{figure}[!t]
    \centering
    \captionsetup{justification=centering}
    \begin{subfigure}{.45\textwidth}
        \centering
        \begin{tikzpicture}[>=stealth',auto,node distance=2cm,main node/.style={minimum size=0.8cm,font=\sffamily\Large\bfseries},scale=1,transform shape]
        \node[main node]   (U) at  (-1.5,-1.1)   {$U$};
        \node[main node]   (X)  at  (0,0)          {$X$};
        \node[main node]   (Y)  at  (2,0)          {$Y$};
        \node[main node]   (Z)  at  (2,-2)      {$Z$};
        \node[main node]   (W) at  (0,-2)      {$W$};
                
        \draw[->]   (U) edge (X);
        \draw[->]   (U) edge (W);
        \draw[->]   (X) edge (Y);
        \draw[<-]   (Y) edge (Z);
        \draw[->]   (Z) edge (W);
        \end{tikzpicture}
        \caption{Unknown DAG}
        \label{fig:ex-no-entner-a}
    \end{subfigure}
    \vrule
    \begin{subfigure}{.45\textwidth}
        \centering
        \begin{tikzpicture}[>=stealth',auto,node distance=2cm,main node/.style={minimum size=0.8cm,font=\sffamily\Large\bfseries},scale=0.9,transform shape]
        \node[main node]   (X) at  (0,0)       {$X$};
        \node[main node]   (Y) at  (2,0)       {$Y$};
        \node[main node]   (Z) at  (2,-2)   {$Z$};
        \node[main node]   (W) at  (0,-2)  {$W$};
                
        \draw[->]   (X) edge (Y);
        \draw[<->]   (W) edge (X);
        \draw[o->]   (Z) edge (W);
        \draw[o->]   (Z) edge (Y);
        \end{tikzpicture}
        \caption{Known PAG}
        \label{fig:ex-no-entner-b}
    \end{subfigure}
    \caption{Graphs used in Examples \ref{ex:no-entner1}-\ref{ex:efficient1}}
    \label{fig:ex-no-entner}
\end{figure}

\begin{example}[Limitations of R1 Entner]
\label{ex:no-entner1}
    Consider a causal model that induces the DAG in Figure \ref{fig:ex-no-entner-a}. Suppose the DAG is unknown, but we have data on $\{X,Y,W,Z\}$ and expert knowledge that ${\{W,Z\} < X < Y}$. We want to know the effect of $X$ on $Y$. 
    
    We can learn from the data that $W \notind Y \,|\, Z$ and ${W \ind Y \,|\, Z,X}$, which implies $\{Z\}$ is an adjustment set relative to $(X,Y)$ by Theorem \ref{thm:r1-entner}. However, there are two adjustment sets Theorem \ref{thm:r1-entner} cannot find that we can find by building a graph from the data. To see this, let the PAG in Figure \ref{fig:ex-no-entner-b} represent all the in/dependencies we can learn from the data with the addition of our expert knowledge.%
    \footnote{Venkateswaran and Perkovi\'c \cite{venkateswaran2024towards} refer to \ref{fig:ex-no-entner-b} as a \textit{restricted essential ancestral graph}.} 
    Using graphical criteria from prior research (see Theorem \ref{thm:as-graphical-pags} in Supp. \ref{supp:prelims}), we can show $\emptyset$ and $\{W,Z\}$ are adjustment sets relative to $(X,Y)$.
\end{example}

%------------------------------------
%  C-EQUIVALENCE
%------------------------------------

\subsection{An Extension}
\label{sec:c-equiv}

To find adjustment sets like those seen in Example \ref{ex:no-entner1}, we note that one can extend Theorem \ref{thm:r1-entner} using the notion of \textit{confounding equivalence} or \textit{c-equivalence} \citep{pearl2014confounding} shown in the definition below.
\begin{definition}
\label{def:c-equiv}
{\normalfont (\textbf{c-equivalence})}
    Let $\mb{X}$, $\mb{Y}$, $\mb{Z}$, and $\mb{T}$ be pairwise disjoint sets of random variables with joint density $f$. Then $\mb{Z}$ and $\mb{T}$ are \textit{c-equivalent} relative to $(\mb{X}, \mb{Y})$ if
    \[ 
    \int f(\mb{y} | \mb{x}, \mb{z}) f(\mb{z}) \diff \mb{z}
    	= \int f(\mb{y} | \mb{x}, \mb{t}) f(\mb{t}) \diff \mb{t}.
    \]
\end{definition}
\citeauthor{pearl2014confounding} use this definition to find sets that, when used for adjustment, produce the same asymptotic bias for estimating a causal effect. For our purposes, we note that if $\mb{T}$ is c-equivalent to an adjustment set relative to $(\mb{X}, \mb{Y})$, then $\mb{T}$ is also an adjustment set relative to $(\mb{X}, \mb{Y})$. 

In Theorem \ref{thm:c-equiv} below, we provide sufficient criteria from prior research for two sets of variables to be c-equivalent. These criteria do not require knowledge of a causal graph, and this will allow us to extend Theorem \ref{thm:r1-entner}.

We note that while Theorem \ref{thm:c-equiv}, as stated, mirrors Corollary 1 of Pearl and Paz \cite{pearl2014confounding}, its conditions can be found throughout prior research (e.g., \citep{greenland1999causal, kuroki2003covariate, kuroki2004selection, de2011covariate}), and across the literature, researchers have used these conditions for similar purposes. We will revisit these conditions in Section \ref{sec:rationale} in discussing the statistical efficiency of adjustment-based estimators.

\begin{theorem}
\label{thm:c-equiv}
{\normalfont (\textbf{Probabilistic Criteria for c-equivalence})}
    Let $\mb{X}$, $\mb{Y}$, and $\mb{Z} \cup \mb{T}$ be pairwise disjoint sets of random variables. Then $\mb{Z}$ and $\mb{T}$ are \textit{c-equivalent} relative to $(\mb{X}, \mb{Y})$ if either of the following hold: 
    \begin{enumerate}[label=(\roman*)]
    	\item $\mb{X} \ind \mb{Z} \,\,|\,\, \mb{T}$ \,\,and\,\, $\mb{Y} \ind \mb{T} \,\,|\,\, \mb{Z} \cup \mb{X}$ \label{thm:c-equiv-i}
    	\item $\mb{X} \ind \mb{T} \,\,|\,\, \mb{Z}$ \,\,and\,\, $\mb{Y} \ind \mb{Z} \,\,|\,\, \mb{T} \cup \mb{X}$. \label{thm:c-equiv-ii}
    \end{enumerate}
\end{theorem}

We use Theorem \ref{thm:c-equiv} to extend the work of EHS \cite{entner2013data} in the following way. When the causal effect of $X$ on $Y$ is identifiable, Theorem \ref{thm:r1-entner} will find at least one adjustment set $\mb{Z}$. Then we can search for a set that is c-equivalent to $\mb{Z}$ relative to $(X,Y)$ using Theorem \ref{thm:c-equiv}. By definition, any such set will also be an adjustment set relative to $(X,Y)$. We show in Example \ref{ex:no-entner2} that this procedure can identify adjustment sets that Theorem \ref{thm:r1-entner} cannot.

%----------------------------
% Example - No Entner 2
%----------------------------

\begin{example}[Adjustment via c-equivalence]
\label{ex:no-entner2}
    Reconsider Example \ref{ex:no-entner1}, where Theorem \ref{thm:r1-entner} could not find the adjustment sets $\emptyset$ and $\{W,Z\}$. Then turn to consider Theorem \ref{thm:c-equiv}. 

    From the data, we can learn that ${X \ind Z}$. And trivially ${Y \ind \emptyset \,|\, Z, X}$. Thus by \ref{thm:c-equiv-i}, $\emptyset$ is c-equivalent to $\{Z\}$ and an adjustment set relative to $(X,Y)$. We already learned from the data that ${Y \ind W \,|\, Z, X}$. And trivially ${X \ind Z \,|\, W, Z}$. Thus by \ref{thm:c-equiv-i}, $\{W, Z\}$ is c-equivalent to $\{Z\}$ and an adjustment set relative to $(X,Y)$. 
\end{example}

%------------------------------------
%  RATIONALE
%------------------------------------

\subsection{Rationale}
\label{sec:rationale}

At first glance, it may seem unimportant to have a choice of sets to use for covariate adjustment. EHS \cite{entner2013data} already have a data-driven method of finding one adjustment set when the causal effect of a single treatment is identifiable, and every adjustment set can be used to construct an unbiased estimator of the causal effect -- given appropriate parametric assumptions, or in a discrete setting, given sufficient data. However, estimators constructed using different adjustment sets may have different statistical properties, such as their asymptotic variance.

Recent research considers adjustment sets that lead to asymptotically efficient estimators of a causal effect -- called \textit{efficient adjustment sets}. Broadly, this research takes two paths: (1) exploring the asymptotic variance of an estimator of the causal effect under assumptions of linearity \citep{henckel2022graphical, witte2020efficient, guo2022efficient, colnet2024re}, or (2) exploring the variance of the influence function of an asymptotically linear estimator of the causal effect in a semi-parametric setting with discrete treatment \citep{smucler2022efficient}. 

In an effort to obtain more efficient estimators, both research paths use the conditions of Theorem \ref{thm:c-equiv} as guidance for adding or removing variables from an adjustment set. We provide a result for evaluating such variables below.

\begin{lemma}
\label{lem:adjustment-vars}
{\normalfont (\textbf{Precision and Overadjustment Variables}, cf. Lemmas 4-5 of \cite{smucler2022efficient}, Theorem 1 of \cite{henckel2022graphical})}
    Let $\mb{X}$, $\{Y\}$, $\mb{Z}$, and $\mb{T}$ be pairwise disjoint sets of random variables in a causal model, where both $\mb{Z}$ and $\mb{Z} \cup \mb{T}$ are adjustment sets relative to $(\mb{X},Y)$. Then $\mb{T}$ are precision variables and $\mb{Z} \cup \mb{T}$ is a more efficient adjustment set compared to $\mb{Z}$ if
    \begin{enumerate}[label=(\roman*)]
    	\item ${\mb{X} \ind \mb{T} \,\,|\,\, \mb{Z}}$.
    \end{enumerate}
$\mb{T}$ are overadjustment variables and $\mb{Z} \cup \mb{T}$ is a less efficient adjustment set  compared to $\mb{Z}$ if
    \begin{enumerate}[label=(\roman*), start=2]
    	\item ${Y \ind \mb{T} \,\,|\,\, \mb{Z} \cup \mb{X}}$.
    \end{enumerate}  
\end{lemma}

Note that in the result above, $\mb{Z \cup T}$ being more (or less) efficient refers to the asymptotic properties of a causal effect estimator that relies on adjustment through $\mb{Z} \cup \mb{T}$. In Example \ref{ex:efficient1} below, we use Lemma \ref{lem:adjustment-vars} to show that Theorem \ref{thm:r1-entner} may find an adjustment set that leads to asymptotically efficient estimation of the causal effect. But Example \ref{ex:efficient2} shows this is not always the case.

%---------------------------------
%   Example - Efficient 1
%---------------------------------

\begin{example}[Efficient Adjustment via R1 Entner]
\label{ex:efficient1}
    Reconsider Examples \ref{ex:no-entner1}-\ref{ex:no-entner2}, where Theorem \ref{thm:r1-entner} found $\{Z\}$ and Theorem \ref{thm:c-equiv} found $\emptyset$ and $\{W,Z\}$ as adjustment sets. We want to know which set leads to an asymptotically efficient estimator of the causal effect. By Lemma \ref{lem:adjustment-vars}, we see that $\{Z\}$ is more efficient than $\emptyset$, since ${X \ind Z}$. Similarly, $\{Z,W\}$ is less efficient than $\{Z\}$, since ${Y \ind W \,|\, Z,X}$. Thus, $\{Z\}$ is the most efficient adjustment set -- and one that Theorem \ref{thm:r1-entner} was able to find.
\end{example}

%---------------------------------
%   Example - Efficient 2
%---------------------------------

\begin{figure}
    \centering
    \captionsetup{justification=centering}
    \begin{tikzpicture}[>=stealth',auto,node distance=2cm,main node/.style={minimum size=0.8cm,font=\sffamily\Large\bfseries},scale=0.9,transform shape]
        \node[main node]   (U1) at  (-1.5,-1)   {$U_1$};
        \node[main node]   (U2) at  (1.5,-1)   {$U_2$};
        \node[main node]   (U3) at  (4.5,-1)   {$U_3$};

        \node[main node]   (X) at  (0,0)   {$X$};
        \node[main node]   (Y) at  (3,0)   {$Y$};
        \node[main node]   (Z) at  (3,-2)   {$Z$};
        \node[main node]   (W) at  (0,-2)   {$W$};

        \draw[->]   (U1) edge (X);
        \draw[->]   (U1) edge (W);
        \draw[->]   (U2) edge (W);
        \draw[->]   (U2) edge (Z);   
        \draw[->]   (U3) edge (Y);
        \draw[->]   (U3) edge (Z);     
        \draw[->]   (X) edge (Y);
    \end{tikzpicture}
    \caption{Unknown DAG used in Example \ref{ex:efficient2}}
    \label{fig:ex-efficient2}
\end{figure}

\begin{example}[Inefficient Adjustment via R1 Entner]
\label{ex:efficient2}
    Consider a causal model that induces the DAG in Figure \ref{fig:ex-efficient2}. Suppose the DAG is unknown, but we have data on $\{X,Y,W,Z\}$ and expert knowledge that ${\{W,Z\} < X < Y}$. We want to know the effect of $X$ on $Y$. 
    
    By Theorem \ref{thm:r1-entner}, $\emptyset$ is an adjustment set relative to $(X,Y)$, since we can learn from the data that ${W \notind Y}$ and ${W \ind Y \,|\, X}$. Theorem \ref{thm:r1-entner} finds no further sets, but Theorem \ref{thm:c-equiv} finds $\{Z\}$ and $\{W\}$ since we can learn from the data that ${X \ind Z}$ and ${Y \ind W \,|\, X}$. By Lemma \ref{lem:adjustment-vars}, we see that $\{Z\}$ adds precision and $\{W\}$ overadjusts compared to $\emptyset$, since ${X \ind Z}$ and ${Y \ind W \,|\, X}$. Thus $\{Z\}$ is again the most efficient adjustment set -- but one that Theorem \ref{thm:r1-entner} was unable to find.
\end{example}

%%%%%%%%%%%%%%%%%%%%%%%%%%%%%%%%%%%%
%      RESULTS - R1                %
%%%%%%%%%%%%%%%%%%%%%%%%%%%%%%%%%%%%

\section{Extension to Multiple Treatments}
\label{sec:r1}

In this section, we provide two paths (Theorems \ref{thm:r1-build} and \ref{thm:r1-combine}) to finding adjustment sets that rely on dependencies and independencies in the observed data directly. Both paths consider a setting with multiple treatments and thus, extend the work of EHS \cite{entner2013data}. As in Theorem \ref{thm:r1-entner}, our methods require that treatments cannot be causal ancestors of observed variables in the model, a condition satisfied when covariates are pre-treatment.

The adjustment set we offer in Theorem \ref{thm:r1-build} is constructed from the ground up. That is, a researcher must find an adjustment set for one element of a set of treatments and then build, element by element, to an adjustment set for all treatments. The adjustment set we offer in Theorem \ref{thm:r1-combine} is constructed by carefully combining adjustment sets for each element in a set of treatments. Notably, our latter method relies on the notion of c-equivalence that we saw in Section \ref{sec:c-equiv}.

%----------------------------
%   R1 Build
%----------------------------

\subsection{Building on Adjustment Sets}
\label{sec:r1-build}

Below we present our first path for extending Theorem \ref{thm:r1-entner} to multiple treatments. Example \ref{ex:build} illustrates its use.

\begin{theorem}
\label{thm:r1-build}
{\normalfont (\textbf{R1 Build})}
    Let ${\mb{X} = \{X_1, \dots, X_k\}}$, $k \ge 1$; $\{Y\}$; and $\mb{W}$ be pairwise disjoint sets of observed random variables in a causal model, and for $i \in \{1, \dots, k\}$, define $\{X_1, \dots, X_i\}^{-i} = \{X_1, \dots, X_i\} \setminus \{X_i\}$.

    Suppose ${\mb{W} < X_1 < \dots < X_k < Y}$ is a causal ordering consistent with the model. If there exist ${W_1, \dots, W_k \in \mb{W}}$ and ${\mb{Z} \subseteq \mb{W} \setminus \{W_1, \dots, W_k \}}$ such that for all $i$,
        \begin{enumerate}[label=(\roman*)]
            \item $W_i \notind Y \,\,|\,\, \mb{Z} \cup  \{X_1, \dots, X_i\}^{-i}$ and \label{thm:r1-build-i}
            \item $W_i \ind Y \,\,|\,\, \mb{Z} \cup \{X_1, \dots, X_i\}$, \label{thm:r1-build-ii}
        \end{enumerate}
    then $\mb{X}$ has a causal effect on $Y$ that is identifiable through the adjustment set $\mb{Z}$.
\end{theorem}

\begin{proofsketch}
    The proof for Theorem \ref{thm:r1-build} can be found in Supp. \ref{supp:r1-build}, but we provide an outline here for intuition. To see that $\mb{X}$ causes $Y$, note that \ref{thm:r1-build-i}-\ref{thm:r1-build-ii} require a path $p_i$ from $W_i$ to $Y$ that is open given $\mb{Z} \cup \{X_1, \dots, X_i\}^{-i}$ and contains $X_i$ as a non-collider. This combined with the causal ordering requires $p_i$ to end $X_i \to \dots \to Y$. To show $\mb{Z}$ is an adjustment set, we only need $\mb{Z} \cup \mb{X^{-i}}$ to block all back-door paths from $X_i$ to $Y$. We prove this holds for ${i=k}$ and proceed by induction. For contradiction in the base case, we assume a back-door path $q_k$ from $X_k$ to $Y$ that is open given $\mb{Z} \cup \mb{X^{-k}}$. Then we define $r_k = p_k(W_k, A) \oplus q_k(A,Y)$ for the earliest shared node $A$. Showing $r_k$ is open given $\mb{Z} \cup \mb{X}$ contradicts condition \ref{thm:r1-build-ii}. This holds for $p_k(W_k, A)$ and $q_k(A,Y)$ by definition. We complete the base case by showing it holds for $r_k$: when $A=X_k$, $A \in \mb{Z} \cup \mb{X^{-k}}$, and $A \notin \mb{Z} \cup \mb{X}$. The induction step follows a similar argument, where we solve two additional issues with the induction assumption.
\end{proofsketch}

%----------------------------
%   Example - Build
%----------------------------

\begin{figure}
    \centering
    \captionsetup{justification=centering}
    \begin{tikzpicture}[>=stealth',auto,node distance=2cm,main node/.style={minimum size=0.8cm,font=\sffamily\Large\bfseries},scale=0.9,transform shape]
        \node[main node]   (X1)  at  (0,0)            {$X_1$};
        \node[main node]   (X2)  at  (3,0)            {$X_2$};
        \node[main node]   (Y)    at  (6,0)            {$Y$};
        \node[main node]   (W1) at  (0,-1.35)      {$W_1$};
        \node[main node]   (W2) at  (3,-1.35)       {$W_2$};
        \node[main node]   (Z1)  at  (1.5,1.25)     {$Z_1$};
        \node[main node]   (Z2)  at  (4.5,1.25)     {$Z_2$};
                
        \draw[->]	(W1) edge (X1);
        \draw[->]	(X1) edge (X2);
        \draw[->]	(X2) edge (Y);
        \draw[->]       (X1) edge[bend right=22] node [left] {} (Y);
        \draw[->]	(Z1) edge (X1);
        \draw[->]	(Z1) edge (X2);
        \draw[->]	(Z1) edge (Y);
        \draw[->]	(W2) edge (X2);
        \draw[->]	(Z2) edge (X2);
        \draw[->]	(Z2) edge (Y);
    \end{tikzpicture}
    \caption{Unknown DAG used in Example \ref{ex:build}}
    \label{fig:ex-build}
\end{figure}

\begin{example}[Adjustment via R1 Build]
\label{ex:build}
    Consider a causal model that induces the DAG in Figure \ref{fig:ex-build}. Suppose the DAG is unknown, but we have data on every variable and expert knowledge that $\{W_1,W_2,Z_1,Z_2\}<X_1<X_2<Y$. To find the effect of $\mb{X}$ on $Y$, note that by Theorem \ref{thm:r1-build}, ${\mb{Z}:=\{Z_1,Z_2\}}$ is an adjustment set relative to $(\mb{X}, Y)$, since we can learn from data that 
    ${W_1 \notind Y \,|\, \mb{Z}}$ and
    ${W_1 \ind Y \,|\, \mb{Z} \cup \{X_1\}}$ as well as
    ${W_2 \notind Y \,|\, \mb{Z} \cup \{X_1\}}$ and 
    ${W_2 \ind Y \,|\, \mb{Z} \cup \{X_1, X_2\}}$.
\end{example}

Theorem \ref{thm:r1-build} is especially useful in settings where Theorem \ref{thm:r1-entner} has already found an adjustment set for a causal effect on a single treatment, and a researcher would like to consider the addition of further treatments. But this method, while intuitive, has its limitations. We showcase this in the example below as motivation for our final extension of Theorem \ref{thm:r1-entner}.

%-------------------------------------
%   Example - Build Limits 1
%-------------------------------------

\begin{figure}[!t]
    \centering
    \captionsetup{justification=centering}
    \begin{subfigure}{.45\textwidth}
        \centering
        \begin{tikzpicture}[>=stealth',auto,node distance=2cm,main node/.style={minimum size=0.8cm,font=\sffamily\Large\bfseries},scale=0.9,transform shape]
   	        \node[main node]   (X1) at  (0,0)   	{$X_1$};
            \node[main node]   (X2) at  (2,0)   	{$X_2$};
            \node[main node]   (Y)   at  (3.75,0)   	{$Y$};
            \node[main node]   (U)  at  (-1.35,.75)	{$U$};
            \node[main node]   (W)  at  (0,1.5)  	{$W$};
            \node[main node]   (Z)   at  (2,1.5)   	{$Z$};

            \draw[->]   	(X1) edge (X2);
            \draw[->]   	(X2) edge (Y);
            \draw[->]		(X1) edge[bend right=20] node [left] {} (Y);
            \draw[->]		(U)  edge (W);
            \draw[->]		(U)  edge (X1);
   	        \draw[->]   	(Z)   edge (X1);
            \draw[->]   	(Z)   edge (X2);
    \end{tikzpicture}
    \caption{Unknown DAG}
    \label{fig:ex-build-limits-a}
    \end{subfigure}
    \vrule
    \begin{subfigure}{.45\textwidth}
        \centering
        \begin{tikzpicture}[>=stealth',auto,node distance=2cm,main node/.style={minimum size=0.8cm,font=\sffamily\Large\bfseries},scale=0.9,transform shape]
   	        \node[main node]   (X1) at  (0,0)   	{$X_1$};
            \node[main node]   (X2) at  (2,0)   	{$X_2$};
            \node[main node]   (Y)   at  (4,0)   	{$Y$};
            \node[main node]   (W)  at  (0,1.5)  	{$W$};
            \node[main node]   (Z)   at  (2,1.5)   	{$Z$};

            \draw[->]   	(X1) edge (X2);
            \draw[->]   	(X2) edge (Y);
            \draw[o->]		(W)  edge (X1);
            \draw[o->]   	(Z)   edge (X1);
            \draw[->]   	(Z)   edge (X2);
            \draw[->]	(X1) edge[bend right=20] node [left] {} (Y);
    \end{tikzpicture}
    \caption{Known PAG}
    \label{fig:ex-build-limits-b}
    \end{subfigure}
\caption{Graphs used in Examples \ref{ex:build-limits1}-\ref{ex:build-limits2}}
\label{fig:ex-build-limits}
\end{figure}

\begin{example}[Limitations of R1 Build]
\label{ex:build-limits1}
    Consider a causal model that induces the DAG in Figure \ref{fig:ex-build-limits-a}. Suppose the DAG is unknown, but we have data on $\{X_1,X_2, W,Y,Z\}$ and expert knowledge that $\{W,Z\}<X_1<X_2<Y$. We attempt to find the effect of $\mb{X}$ on $Y$ using Theorem \ref{thm:r1-build}. To fulfill \ref{thm:r1-build-i}-\ref{thm:r1-build-ii}, we must set ${W_1=W}$ and ${\mb{Z}=\{Z\}}$. But then there is no $W_2$ that fulfills \ref{thm:r1-build-i}-\ref{thm:r1-build-ii}. Thus, we cannot use Theorem \ref{thm:r1-build} to find an adjustment set. 

    However, we can find two adjustment sets relative to $(\mb{X}, Y)$ by building a graph from the data. To see this, let the PAG in Figure \ref{fig:ex-build-limits-b} represent all the in/dependencies we can learn from the data with the addition of our expert knowledge. Using graphical criteria from prior research (see Theorem \ref{thm:as-graphical-pags} in Supp. \ref{supp:prelims}), we can show that $\emptyset$, $\{Z\}$, and $\{W,Z\}$ are adjustment sets relative to $(\mb{X},Y)$.
\end{example}

%----------------------------
%   R1 Combine
%---------------------------- 

\subsection{Combining Adjustment Sets}
\label{sec:r1-combine}

Below we present our second path for extending Theorem \ref{thm:r1-entner} to multiple treatments. Informally, this method constructs an adjustment set for the full set of treatments by combining adjustment sets for the individual treatments -- after first removing extraneous variables. We formalize this notion in the definition below before providing our result.

\begin{definition}
\label{def:minimal-adjust}
{\normalfont (\textbf{Minimal Adjustment Set})}
    A set $\mb{Z}$ is a \textit{minimal adjustment set} relative to $(\mb{X}, \mb{Y})$ if $\mb{Z}$ is an adjustment set relative to $(\mb{X}, \mb{Y})$ and no proper subset of $\mb{Z}$ is an adjustment set relative to $(\mb{X}, \mb{Y})$.
\end{definition}

\begin{theorem}
\label{thm:r1-combine}
{\normalfont (\textbf{R1 Combine})}
    Let $\mb{X} = \{ X_1, \dots, X_k\}$, $k \ge 1$; $\{Y\}$; and $\mb{W}$ be pairwise disjoint sets of observed random variables in a causal model, and for $i \in \{1, \dots, k\}$, define $\mb{X^N_i}$ to be the variables in $\mb{X}$ that are not causal descendants of $X_i$.

    Suppose ${\mb{W} < \mb{X} < Y}$ is a causal ordering consistent with the model. If there exist $W_1, \dots, W_k \in \mb{W}$ and $\mb{T_i} \subseteq \big[ \mb{W} \setminus \{W_i \} \big] \cup \mb{X^N_i}$ such that for all $i$,
    \begin{enumerate}[label=(\roman*)]
            \item $W_i \notind Y \,\,|\,\, \mb{T_i}$ \, and \label{thm:r1-combine-i}
            \item $W_i \ind Y \,\,|\,\, \mb{T_i} \cup \{X_i\}$, \label{thm:r1-combine-ii}
    \end{enumerate}
    then $\mb{X}$ has a causal effect on $Y$ that is identifiable through the adjustment set $\mb{Z} := \cup_{i=1}^k \mb{Z_i} \setminus \mb{X}$, where $\mb{Z_i}$ is any minimal adjustment set relative to $(X_i, Y)$ such that $\mb{Z_i} \subseteq \mb{T_i}$.
\end{theorem}

\begin{proofsketch}
    The proof for Theorem \ref{thm:r1-combine} can be found in Supp. \ref{supp:r1-combine}, but we provide an outline here for intuition. Note that \ref{thm:r1-combine-i}, \ref{thm:r1-combine-ii}, and Theorem \ref{thm:r1-entner} imply $X_i$ causes $Y$ and $\mb{T_i}$ is an adjustment set relative to $(X_i,Y)$. Thus if $\mb{T_i}$ exists, then the reduced adjustment set $\mb{Z_i} \subseteq \mb{T_i}$ is guaranteed. To show that $\mb{Z}$ is an adjustment set relative to $(\mb{X},Y)$, we only need $\mb{Z} \cup \mb{X^{-i}}$ to block all back-door paths from $X_i$ to $Y$, where we define $\mb{X^{\text{-}i}} = \mb{X} \setminus \{X_i\}$. Without loss of generality, we show this holds for ${i=1}$. For contradiction, we assume a back-door path $q$ from $X_1$ to $Y$ that is open given $\mb{Z} \cup \mb{X^{-1}}$. Since $\mb{Z_1}$ is an adjustment set relative to $(X_1,Y)$, $q$ must have a collider that is a causal ancestor of $\mb{Z} \cup \mb{X^{-1}}$ but not $\mb{Z_1}$. Using the minimality of each $\mb{Z_i}$, we show that all such colliders $C_1, \dots, C_\ell$ must have directed paths $t_1, \dots, t_\ell$ to $Y$, which we use to define a final back-door path $u$ from $X_1$ to $Y$ that is open given $\mb{Z_1}$. For example, when there is no directed path from $\{C_1, \dots, C_\ell\}$ to $X_1$, then $u := q(X_1, C_1) \oplus t_1$. This path contradicts that $\mb{Z_1}$ is an adjustment set relative to $(X_1,Y)$.
\end{proofsketch}

Implementing Theorem \ref{thm:r1-combine} requires checking if each $\mb{T_i}$ is a minimal adjustment set, and if not, then finding such a set $\mb{Z_i}$. On its face, this involves knowledge of either the underlying graph or the underlying joint density of observed variables. However, a key appeal of the rules of EHS \cite{entner2013data} -- that we aim to replicate -- is the lack of reliance on graphical criteria. To resolve this discrepancy, we present the lemma below, which provides a route for finding a minimal adjustment set through the testing of in/dependencies among observed variables in the data directly.

% ------- MINIMALITY CRITERIA ------- %

\begin{lemma}[Probabilistic Criteria for Minimality]
\label{lem:minimal-criteria} 
    Let $\mb{T}$ be an adjustment set relative to $(X, Y)$ in a causal model where $\mb{T} < X$. Then $\mb{T}$ is a minimal adjustment set relative to $(X,Y)$ if and only if for all $T \in \mb{T}$,
    \begin{enumerate}[label=(\roman*)]
        \item $X \notind T \,\,|\,\, \mb{T} \setminus \{T\}$, and \label{lem:minimal-criteria-ai}
        \item $Y \notind T \,\,|\,\, \big[ \mb{T} \setminus \{T\} \big] \cup \{X\}$. \label{lem:minimal-criteria-aii}
    \end{enumerate}
    Otherwise, $\mb{Z} \subset \mb{T}$ is a minimal adjustment set relative to $(X,Y)$ if for all $Z \in \mb{Z}$,
    \begin{enumerate}[label=(\roman*), start=3]
        \item $X \notind Z \,\,|\,\, \mb{Z} \setminus \{Z\}$, \label{lem:minimal-criteria-bi}
        \item $Y \notind Z \,\,|\,\, \big[ \mb{Z} \setminus \{Z\} \big] \cup \{X\}$, and \label{lem:minimal-criteria-bii}
        \item $Y \ind \mb{T} \,\,|\,\, \mb{Z} \cup \{X\}$ \,\,\,or\,\,\, $X \ind \mb{T} \,\,|\,\, \mb{Z}$. \label{lem:minimal-criteria-biii}
    \end{enumerate}
\end{lemma}

\begin{proofsketch}
    The proof for Lemma \ref{lem:minimal-criteria} can be found in Supp. \ref{supp:r1-combine}, but for intuition, we note that \ref{lem:minimal-criteria-ai}-\ref{lem:minimal-criteria-aii} require a path from $X$ to $T$ that is open given $\mb{T} \setminus \{T\}$, and a path from $T$ to $Y$ that is open given $\{X\} \cup \mb{T} \setminus \{T\}$. Informally, combining these paths offers a non-causal path from $X$ to $Y$ that is open given $\mb{T} \setminus \{T\}$. Thus, $\mb{T}$ is \textit{elementwise minimal}, meaning $\mb{T} \setminus \{T\}$ is not an adjustment set for any $T \in \mb{T}$, which we show implies minimality. When \ref{lem:minimal-criteria-ai}-\ref{lem:minimal-criteria-aii} do not hold, \ref{lem:minimal-criteria-biii} and Theorem \ref{thm:c-equiv} show $\mb{Z} \subset \mb{T}$ is an adjustment set, and the proof for the minimality of $\mb{Z}$ follows similarly from \ref{lem:minimal-criteria-bi}-\ref{lem:minimal-criteria-bii}.
\end{proofsketch}

We explore Theorem \ref{thm:r1-combine} and Lemma \ref{lem:minimal-criteria} in the examples below. Example \ref{ex:build-limits2} provides a straightforward demonstration of these results, and Example \ref{ex:combine-naive1} shows why we require minimality.

%-------------------------------------
%   Example - Build Limits 2
%-------------------------------------

\begin{example}[Adjustment via R1 Combine] 
\label{ex:build-limits2}
    Reconsider the setting of Example \ref{ex:build-limits1}. While Theorem \ref{thm:r1-build} could not find an adjustment set, we show Theorem \ref{thm:r1-combine} will. Let ${W_1=W}$, ${\mb{T_1}=\{Z\}}$, ${W_2=Z}$, and ${\mb{T_2}=\{X_1\}}$. Note that \ref{thm:r1-combine-i} and \ref{thm:r1-combine-ii} hold, since ${W \notind Y \,|\, Z}$ and ${W \ind Y \,|\, Z, X_1}$ as well as ${Z \notind Y \,|\, X_1}$ and ${Z \ind Y \,|\, X_1, X_2}$. By Lemma \ref{lem:minimal-criteria}, we see that $\mb{T_1}$ is a minimal adjustment set relative to $(X_1,Y)$, since ${X_1 \notind Z}$ and ${Y \notind Z \,|\, X_1}$. The analogous claim holds for $\mb{T_2}$, since ${X_2 \notind X_1}$ and ${Y \notind X_1 \,|\, X_2}$. Thus by Theorem \ref{thm:r1-combine}, $\{Z\}$ is an adjustment set relative to $(\mb{X},Y)$.
\end{example}

%-------------------------------------------
%   Example - Naive Combine 1 
%-------------------------------------------

\begin{figure}
    \centering
    \captionsetup{justification=centering}
    \begin{subfigure}{.45\textwidth}
        \centering
        \begin{tikzpicture}[>=stealth',auto,node distance=2cm,main node/.style={minimum size=0.6cm,font=\sffamily\Large\bfseries},scale=0.9,transform shape]
            \node[main node]   (X1)  at  (-.5,0)         {$X_1$};
            \node[main node]   (Y)    at  (2,0)           {$Y$};
            \node[main node]   (X2)  at  (4,0)           {$X_2$};
            \node[main node]   (Z1)  at  (.3,2)          {$Z_1$};
            \node[main node]   (Z2)  at  (3.2,2)           {$Z_2$};
            \node[main node]   (W1) at  (-.5,3)         {$W_1$};
            \node[main node]   (W2) at  (4,3)           {$W_2$};
            \node[main node]   (U1)  at  (-.5,1.5)      {$U_1$};
            \node[main node]   (U2)  at  (1.65,2)      {$U_2$};
            \node[main node]   (U3)  at  (1.15,.95)   {$U_3$};

            \draw[->]     (X1)  edge (Y);
            \draw[<-]     (Y)    edge (X2);
            \draw[<-]     (W1) edge (Z1);
            \draw[->]     (W2) edge (Z2);
            \draw[->]     (W2) edge (X2);
            \draw[->]     (Z2)  edge (Y);
            \draw[->]     (Z2)  edge (X2);
            \draw[->]     (U1) edge (X1);
            \draw[->]     (U1) edge (W1);
            \draw[->]     (U2) edge (Z1);
            \draw[->]     (U2) edge (Z2);
            \draw[->]     (U3) edge (Z1);
            \draw[->]     (U3) edge (Y);
        \end{tikzpicture}
        \caption{Unknown DAG}
        \label{fig:ex-combine-naive1-a}
    \end{subfigure}
    \vrule
    \begin{subfigure}{.45\textwidth}
        \centering
        \begin{tikzpicture}[>=stealth',auto,node distance=2cm,main node/.style={minimum size=0.6cm,font=\sffamily\Large\bfseries},scale=0.9,transform shape]
            \node[main node]   (X1)  at  (0,0)     {$X_1$};
            \node[main node]   (Y)    at  (2,0)     {$Y$};
            \node[main node]   (X2)  at  (4,0)     {$X_2$};
            \node[main node]   (Z1)  at  (.8,1.7)  {$Z_1$};
            \node[main node]   (Z2)  at  (3.1,1.7)  {$Z_2$};
            \node[main node]   (W1) at  (0,3)     {$W_1$};
            \node[main node]   (W2) at  (4,3)     {$W_2$};

            \draw[->]     (X1)  edge (Y);
            \draw[<-]     (Y)    edge (X2);
            \draw[<->]   (W1) edge (X1);
            \draw[<-]     (W1) edge (Z1);
            \draw[o->]   (W2) edge (Z2);
            \draw[->]     (W2) edge (X2);
            \draw[<->]   (Z1)  edge (Z2);
            \draw[<->]   (Z1)  edge (Y);
            \draw[->]     (Z2)  edge (Y);
            \draw[->]     (Z2)  edge (X2);
        \end{tikzpicture}
        \caption{Known PAG}
        \label{fig:ex-combine-naive1-b}
    \end{subfigure}
    \caption{Graphs used in Example \ref{ex:combine-naive1}}
    \label{fig:ex-combine-naive1}  
\end{figure}

\begin{example}[Limitations of Naive Combinations]
\label{ex:combine-naive1}
    Consider a causal model that induces the DAG in Figure \ref{fig:ex-combine-naive1-a}. Suppose the DAG is unknown, but we have data on every variable except $\{U_1,U_2,U_3\}$ and expert knowledge that $\{W_1,W_2,Z_1,Z_2\} < \{X_1,X_2\} <Y$.

    We consider constructing an adjustment set relative to $(\{X_1,X_2\},Y)$ by combining one for $(X_1,Y)$ with one for $(X_2,Y)$ without requiring minimality. By Theorem \ref{thm:r1-entner}, $\{Z_1\}$ and $\{Z_2\}$ are adjustment sets relative to $(X_1,Y)$ and $(X_2,Y)$, respectively, since ${W_1 \notind Y \,|\, Z_1}$; ${W_1 \ind Y \,|\, Z_1, X_1}$; ${W_2 \notind Y \,|\, Z_2}$; and ${W_2 \ind Y \,|\, Z_2, X_2}$. However, we can show that $\{Z_1,Z_2\}$ is not an adjustment set relative to $(\{X_1,X_2\},Y)$.

    To see this, let the PAG in Figure \ref{fig:ex-combine-naive1-b} represent all the in/dependencies we can learn from the data with the addition of our expert knowledge. The claim follows by graphical criteria from prior research (see Theorem \ref{thm:as-graphical-pags} in Supp. \ref{supp:prelims}). Had we required minimality, we would have found that $\{Z_1\}$ is not minimal. (This holds by Lemma \ref{lem:minimal-criteria}, since ${X_1 \ind Z_1}$.) Thus, Theorem \ref{thm:r1-combine} finds $\{Z_2\}$ is an adjustment set relative to $(\{X_1,X_2\},Y)$, which Figure \ref{fig:ex-combine-naive1-b} confirms.
\end{example}

%----------------------------
%   Generalizing Section 3
%---------------------------- 

\subsection{C-Equivalence and Efficiency}

We close this section by showing how to extend Theorems \ref{thm:r1-build} and \ref{thm:r1-combine} using the methods of Section \ref{sec:extension1}.

\begin{example}[Extending R1 Build, R1 Combine]
	Revisit Example \ref{ex:build-limits2}, where Theorem \ref{thm:r1-combine} found $\{Z\}$ as an adjustment set relative to $(\mb{X},Y)$. This is, in fact, the only adjustment set Theorem \ref{thm:r1-combine} finds for this causal effect. However, as in Section \ref{sec:c-equiv}, we can use c-equivalence to find $\emptyset$ and $\{W,Z\}$. (This holds by Theorem \ref{thm:c-equiv}, since ${Y \ind \{W,Z\} \,|\, \mb{X}}$.) Then as in Section \ref{sec:rationale}, we can use Lemma \ref{lem:adjustment-vars} to see that $\{Z\}$ and $\{W,Z\}$ are both less efficient than $\emptyset$, since ${Y \ind \{W,Z\} \,|\, \mb{X}}$. Thus, adjustment via $\emptyset$ may lead to more efficient estimation of the causal effect.
\end{example}

%%%%%%%%%%%%%%%%%%%%%%%%%%%%%%%%%%%%
%           SIMULATIONS            %
%%%%%%%%%%%%%%%%%%%%%%%%%%%%%%%%%%%%

\section{Simulations}
\label{sec:simulations}

In this section, we use simulations to illustrate how our methods perform in settings with multiple treatments. To do this, we simulate data from random DAGs and then attempt to identify an adjustment set using Theorems \ref{thm:r1-build} and \ref{thm:r1-combine} on ``observed'' variables in the generated data. We compare the performance of our data-driven methods against the performance of an existing graphical approach across three metrics: accuracy in identification, accuracy in estimation, and running time. We find that our methods outperform the graphical approach when a treatment effect exists -- by all metrics except running time in high dimensions.

% -------------  DATA GENERATION  -------------%

\subsection{Data Generation}

To generate data, we start by building a random DAG. We assume a causal ordering of its nodes $V_1 < \dots < V_{15}$ and assign parents in the DAG as follows. For each $V_i$, we choose the size $k$ of its parent set uniformly between zero and $\min(i-1,3)$. Then we choose $k$ variables from $\{V_1, \dots, V_{i-1}\}$ at random. In this DAG, we designate
\begin{align*} 
    V_1, \dots, V_5&: \text{ unobserved covariates,} \\
    V_6, \dots, V_{12}&: \text{ observed covariates } \mb{W}, \\
    V_{13},V_{14}&: \text{ observed treatments } X_1,X_2 \text{, and} \\
    V_{15}&: \text{ observed outcome } Y.
\end{align*}
Using this DAG, we consider two characteristics about the model. First, we note if the model contains a treatment effect by checking to see if the DAG contains $X_1 \to Y$ or $X_2 \to Y$. Second, we note if there is an adjustment set relative to $(\{X_1,X_2\},Y)$ that we can learn from observed data generated by the model. To check this, we consider the PAG that represents in/dependencies from such data as well as knowledge that ${\mb{W} < X_1 < X_2 < Y}$. To confirm this PAG contains an adjustment set, we use graphical criteria from prior research (see Theorem \ref{thm:as-graphical-pags} in Supp. \ref{supp:prelims}). These two characteristics are the basis for the settings we consider in our simulations.
\begin{align*} 
    \text{\textbf{Setting 1}}&\text{\textbf{:} Treatment effect exists, adjustment set exists.}\\
    \text{\textbf{Setting 2}}&\text{\textbf{:} Treatment effect exists, no adjustment set.}\\
    \text{\textbf{Setting 3}}&\text{\textbf{:} No treatment effect.}
\end{align*}
We continue generating graphs in this way until we have 100 models that fall into each of the three settings above. From each DAG, we generate three datasets -- one with 100 observations; one with 1,000; and one with 10,000 -- using the following linear Gaussian structural equation model (SEM). We let each random variable be a linear combination of its parents in the DAG $\g[D]$ and an error term $\varepsilon_i \sim N(0, \sigma^2_i)$, where $\{\varepsilon_1, \dots, \varepsilon_{15}\}$ are mutually independent:
\begin{align*}
    V_i \ \longleftarrow \mspace{-25mu} \sum_{V_j \in \Pa(V_i, \g[D])} \mspace{-25mu} b_{ij} V_j + \varepsilon_i.
\end{align*}
We choose each $b_{i j}$ uniformly from $[-1.5,-0.1] \cup [0.1,1.5]$ and each $\sigma^2_i$ uniformly from $[0.5,1]$. This process gives us 100 datasets for each combination of setting and sample size.

% -------------  METHODS  -------------%

\subsection{Method Application}

Using each dataset, we attempt to find an adjustment set for the causal effect of $\{X_1,X_2\}$ on $Y$ by applying our data-driven methods (Theorems \ref{thm:r1-build} and \ref{thm:r1-combine}) to the observed data. To find variables $W_i$, $i \in \{1, \dots, k\}$, and a set $\mb{Z}$ that fulfill the in/dependencies of (i) and (ii) of each theorem, we run a brute-force search over all observed variables. For checking each in/dependency, we run a hypothesis test for non-zero partial correlation using Fisher’s z-transformation and a threshold of $0.05$. Further, we allow both methods to assume ${\mb{W}<X_1<X_2<Y}$.

For comparison, we attempt to find an adjustment set for the same causal effect by applying an existing graphical approach -- the \textit{generalized adjustment criterion} (GAC; see Theorem \ref{thm:as-graphical-pags} in Supp. \ref{supp:prelims}) \cite{perkovic2018complete}. In order to apply this method, we must first learn a causal PAG from the observed data. We do this using a causal discovery algorithm known as {\scshape FciTiers} \citep{andrews2020completeness} -- a version of FCI \citep{spirtes2000causation} where variables cannot be causal descendants of downstream tiers. We run this algorithm using four tiers, based on knowledge that ${\mb{W}<X_1<X_2<Y}$. And we check for conditional independencies using hypothesis tests with a threshold of $0.05$.

% -------------  RESULTS  -------------%

\subsection{Results}

\begin{figure*}
    \centering
    \begin{tabular}{c|c}
        \includegraphics[width = .49\columnwidth]{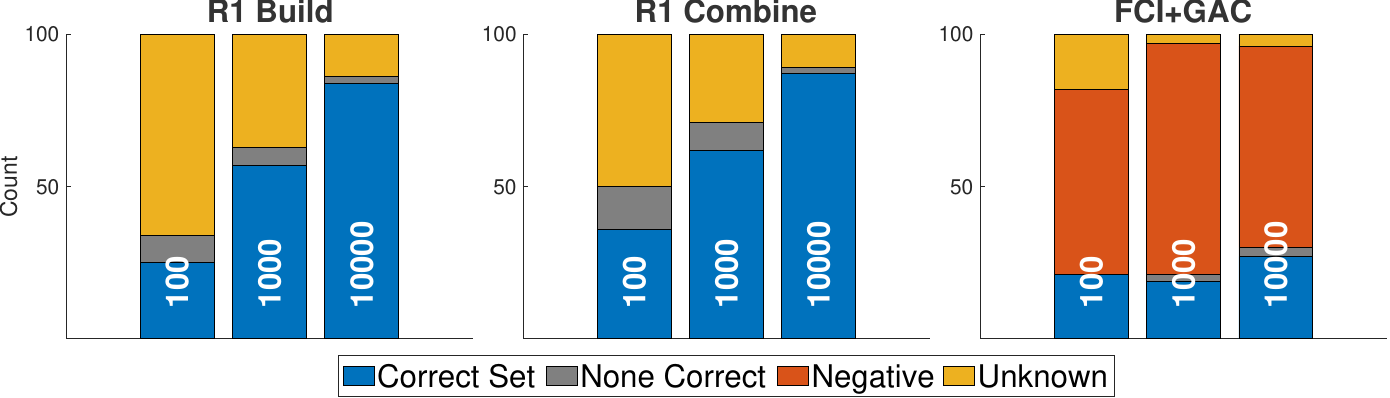}&
        \includegraphics[width = .49\columnwidth]{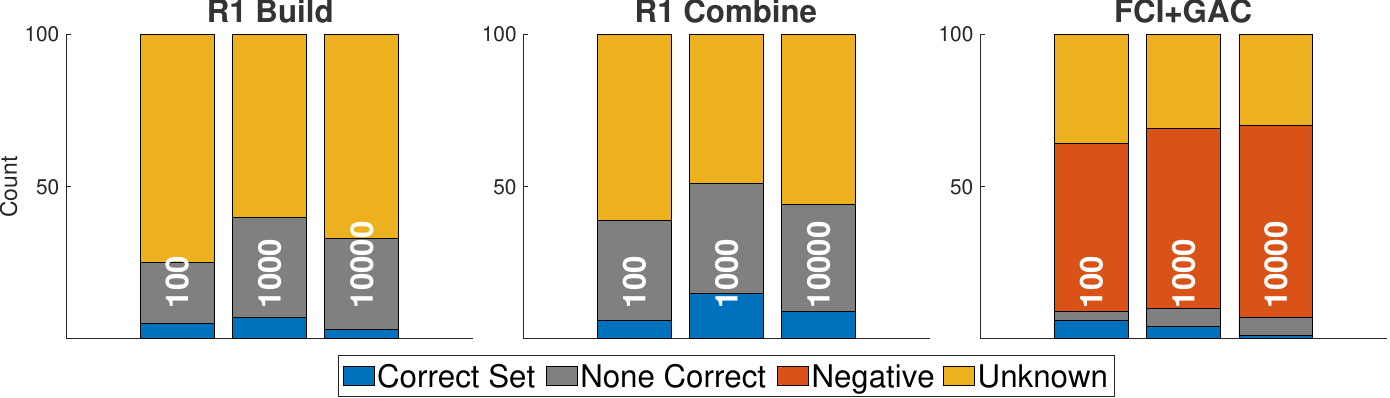}\\
        (a) Setting 1 & (b) Setting 2 \\ %{\color{lightgray} Non-zero Treatment Effect, Adj. Set (PAG)} & {\color{lightgray} Non-zero Treatment Effect, No Adj. Set (PAG)}\\
    \end{tabular}
    \caption{Outcomes of our data-driven methods (R1 Build, R1 Combine) and an existing graphical approach ({FCI + GAC}) on simulated data. Stacked bars show how often a method results in \textit{correct set}, \textit{none correct}, \textit{negative}, or \textit{unknown}.}
\label{fig:sim-conclusions12}
\end{figure*}

To compare how accurately all three methods (R1 Build, R1 Combine, {FCI + GAC}) identify the causal effect, consider the following outcomes.
\begin{itemize}[leftmargin=2.8cm,labelsep=0.3cm]
    \item[\textbf{Correct Set:}] Method concludes $\{X_1, X_2\}$ affects $Y$. At least one adjustment set it finds is correct \\ \hspace*{.5em} in the underlying DAG.
    \vskip .05in

    \item[\textbf{None correct:}] Method concludes $\{X_1, X_2\}$ affects $Y$. No adjustment set it finds is correct in the \\ \hspace*{.5em} underlying DAG.
    \vskip .05in
    
    \item[\textbf{Negative:}] Method concludes $\{X_1, X_2\}$ has no effect on $Y$.
    \vskip .05in

    \item[\textbf{Unknown:}] Method cannot find an adjustment set for the effect of  $\{X_1, X_2\}$ on $Y$.
\end{itemize}
Figure \ref{fig:sim-conclusions12} shows how often each outcome occurs for each method. The bars in each plot correspond to different dataset sizes (100; 1,000; and 10,000). The following outcomes are possible for  all three methods: \textit{correct set}, \textit{none correct}, or \textit{unknown}. {FCI + GAC} can additionally obtain a \textit{negative} outcome, since FCI can learn a PAG that precludes a treatment effect.

Consider Setting 1, shown in Figure \ref{fig:sim-conclusions12}(a), where datasets are generated from models with both a treatment effect and adjustment set (among the observed variables). Successful performance in this setting is when a method finds a correct adjustment set. We see that R1 Build and R1 Combine make many accurate conclusions -- particularly when applied to larger datasets -- since they find a \textit{correct set} in up to 85\% of the datasets. In contrast, {FCI + GAC} makes fewer accurate conclusions, since it finds a \textit{correct set} in under 25\% of the datasets. Further, {FCI + GAC} inaccurately obtains a \textit{negative} outcome in over 60\% of the datasets. We conjecture that our methods outperform {FCI + GAC} in part because they do not attempt to identify the model's entire causal structure and thus avoid errors in hypothesis testing for irrelevant areas of a graph.

Next, consider Setting 2, shown in Figure \ref{fig:sim-conclusions12}(b), where datasets are generated from models with a treatment effect but no adjustment set (among the observed variables). Successful performance in this setting is when a method does not find an adjustment set (i.e., the outcome is \textit{unknown}). Although R1 Build and R1 Combine incorrectly find adjustment sets in 25-50\% of the datasets, {FCI + GAC} incorrectly obtains a \textit{negative} outcome for over 50\% of the datasets.%
\footnote{For clarity in reading Figure \ref{fig:sim-conclusions12}(b), note that in a small percentage of datasets, each method obtains a \textit{correct set} (based on the underlying DAG) even though models in Setting 2 do not have an adjustment set that we can learn from observed data (i.e., in the PAG that represents the model).}

For Setting 3, where datasets are generated from models without a treatment effect, analogous plots can be found in Supp. \ref{supp:simulations}. But we note that {FCI + GAC} easily outperforms R1 Build and R1 Combine in this setting, since our methods cannot obtain a \textit{negative} outcome.

% -------------------------------

\begin{figure*}
    \centering
    \begin{tabular}{c|c}
        \includegraphics[width = .49\columnwidth]{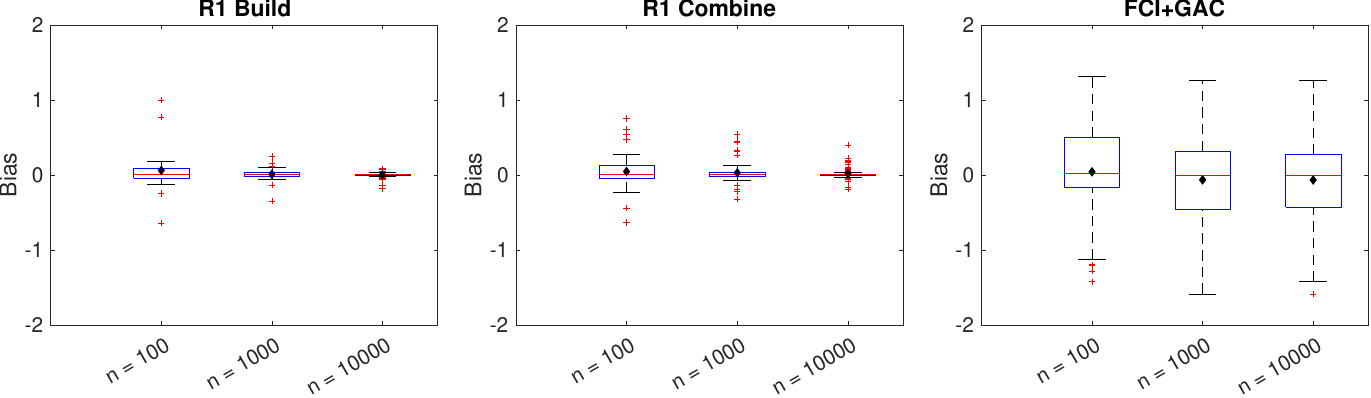}&
        \includegraphics[width = .49\columnwidth]{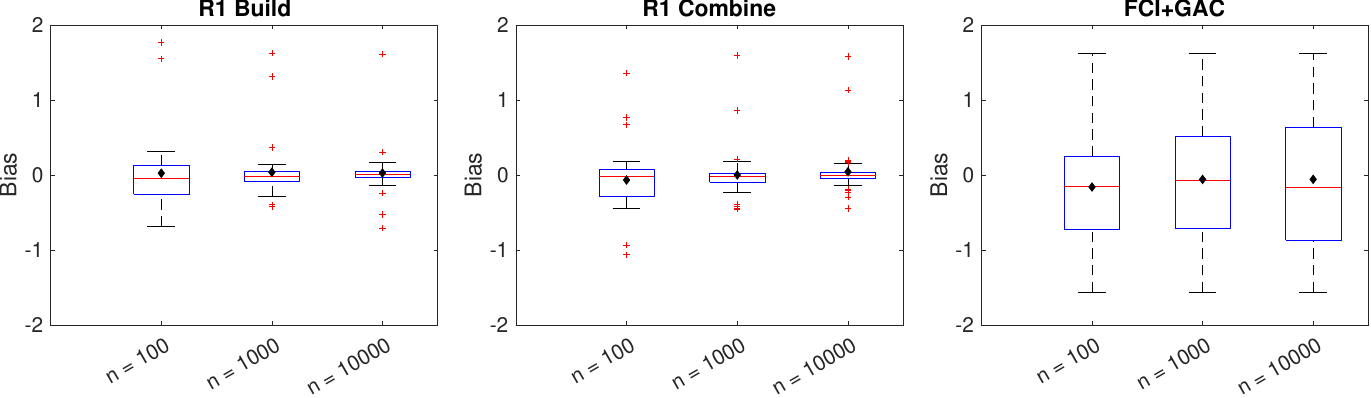} \\
        (a) Setting 1 & (b) Setting 2 \\
    \end{tabular}
    \caption{Differences between estimated and true causal effects. Results for our data-driven methods (R1 Build, R1 Combine) and an existing graphical approach ({FCI + GAC}) when applied to simulated data.}
\label{fig:sim-effectdiffs12}
\end{figure*}

\vskip .1in
We move beyond identification to compare the accuracy of estimating the causal effect based on conclusions from each of the three methods. We do this estimation as follows. If a method finds an adjustment set $\mb{Z}$, we run a linear regression of $Y$ on $\{X_1, X_2\} \cup \mb{Z}$. Then we estimate the causal effect using the sum of estimated coefficients for $X_1$ and $X_2$ from the regression. In the linear setting this corresponds to an estimate of $\mathbb{E}[Y | do(X_1 = x_1+ 1, X_2 = x_2 +1)] - \mathbb{E}[Y| do(X_1 = x_1, X_2 = x_2)]$. Additionally for {FCI + GAC}, if the method obtains a \textit{negative} outcome, then we set the causal effect to zero. If a method obtains an \textit{unknown} outcome, then we do not estimate a causal effect.

Figure \ref{fig:sim-effectdiffs12} shows the differences between these estimated values and the true causal effect across all three methods. Note that estimates based on our methods outperform {FCI + GAC} in both Settings 1 and 2, especially for larger datasets. Between our methods, estimates based on R1 Build perform slightly better than R1 Combine. We conjecture this is due to the larger number of hypothesis tests required to use R1 Combine. Further, estimation improves with increasing sample size -- but only for our methods. 

Though we want to temper these results, because the plots in Figure \ref{fig:sim-effectdiffs12} hide a discrepancy between our methods and the existing graphical approach. Since we do not estimate a causal effect when a method obtains an \textit{unknown} outcome, the estimates for each method in Figure \ref{fig:sim-effectdiffs12} are based on a different subset of the 100 datasets in the simulation. To make this concrete, consider the estimates in Figure \ref{fig:sim-effectdiffs12}(b) from datasets with a sample size of 1,000. Here, R1 Build and R1 Combine find an adjustment set in 39 and 52 datasets, respectively. Thus, we estimate the causal effect based on these methods in 39 and 52 datasets, respectively. But {FCI + GAC} finds either an adjustment set or no treatment effect in 67 datasets, which include some but not all of the 39 and 52 datasets used in estimates based on our methods.%
\footnote{To get a sense of these counts, return to Figure \ref{fig:sim-conclusions12}. We estimate a causal effect when a method obtains the following outcomes: \textit{correct set}, \textit{none correct}, \textit{negative}. These correspond to the blue, gray, and red bars in Figure \ref{fig:sim-conclusions12}.}

For Setting 3, where datasets are generated from models without a treatment effect, analogous plots can be found in Supp. \ref{supp:simulations}. But we note that {FCI + GAC} easily outperforms R1 Build and R1 Combine in this setting, since our methods cannot obtain a \textit{negative} outcome.

% ------------------------------- 

\begin{figure}[t!]
\centering
    \begin{tabular}{c}
        \includegraphics[width = .55\textwidth]{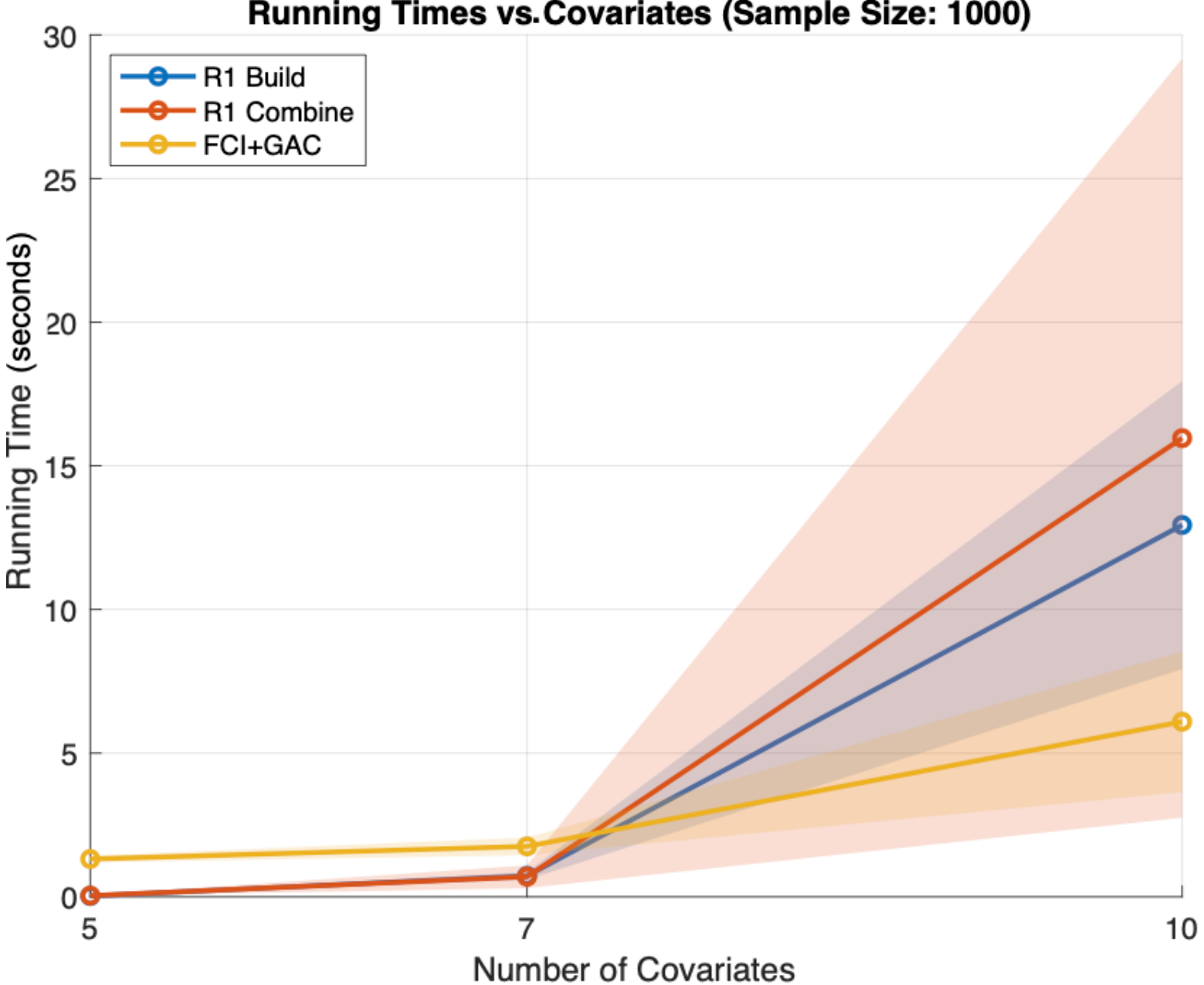}\\
    \end{tabular}
    \caption{Running times for our data-driven methods (R1 Build, R1 Combine) and an existing graphical approach ({FCI + GAC}) when applied to simulated data.}
\label{fig:sim-times}
\end{figure}

\vskip .1in
Finally, we compare all three methods' running times. Figure \ref{fig:sim-times} shows that our data-driven methods perform similarly to {FCI + GAC} on models with five and seven observed covariates. However, {FCI + GAC} outperforms our methods, on average, in models with 10 observed covariates. See Section \ref{sec:discussion} below for further discussion.

%%%%%%%%%%%%%%%%%%%%%%%%%%%%%%%%%%%%
%           DISCUSSION             %
%%%%%%%%%%%%%%%%%%%%%%%%%%%%%%%%%%%%

\section{Discussion}
\label{sec:discussion}

This paper considers causal effect identification through covariate adjustment. While prior research in this area focuses on finding adjustment sets based on criteria from a causal graph \citep{pearl1995causal, shpitser2012validity, maathuis2015generalized, perkovic2018complete}, we consider instead a route that relies on conditional in/dependencies in the observed data directly. This extends the work of EHS \cite{entner2013data}.

We start by reviewing R1 of EHS \cite{entner2013data} and explaining how c-equivalence can extend this rule. We provide a rationale for such an extension in the context of efficient estimation. Then we present our main contributions: Theorems \ref{thm:r1-build} and \ref{thm:r1-combine} (R1 Build and R1 Combine). These data-driven rules parallel R1 of EHS \cite{entner2013data} but in a setting with multiple treatments. Using simulated data, we show that our rules outperform an existing graphical approach in settings where a treatment effect exists.

% --------------  LIMITATION 1  ---------------- %

While the results of our simulations are encouraging, we discuss some limitations of our methods below. We begin by highlighting that Theorems \ref{thm:r1-build} and \ref{thm:r1-combine} are sound but not complete for finding adjustment sets in the presence of multiple treatments. That is, any set these theorems find is a valid adjustment set, but these theorems cannot find every valid adjustment set. To see that the latter holds outside our restricted setting ($\mathbf{W}<\mathbf{X}$), note that there are models where every adjustment set includes descendants of $\mathbf{X}$ (see Examples 7-8 of \cite{perkovic2018complete}). But even within our restricted setting, Theorems \ref{thm:r1-build} and \ref{thm:r1-combine} cannot find an adjustment set in every model. We show this for Theorem \ref{thm:r1-build} in Example \ref{ex:build-limits1}. For Theorem \ref{thm:r1-combine}, see the example below.

\begin{figure}
    \centering
    \captionsetup{justification=centering}
        \begin{tikzpicture}[>=stealth',auto,node distance=2cm,main node/.style={minimum size=0.8cm,font=\sffamily\Large\bfseries},scale=0.9,transform shape]
            \node[main node]   (V3) at  (-2,-1)   {$V_3$};
            \node[main node]   (V2) at  (1.75,-2.5) {$V_2$};
            \node[main node]   (V1) at  (1.75,.5)   {$V_1$};
            \node[main node]   (X1) at  (0,0)   {$X_1$};
            \node[main node]   (Y)  at  (3,-1)  {$Y$};
            \node[main node]   (X2) at  (0,-2)  {$X_2$};

            \draw[o->]  (V3) edge (X1);
            \draw[->]   (V3) edge (X2);
            \draw[o->]  (V2) edge (X2);
            \draw[->]   (V2) edge (Y);
            \draw[o->]  (V1) edge (X1);
            \draw[->]   (V1) edge (Y);
            \draw[->]   (X1) edge (Y);
            \draw[->]   (X2) edge (Y);
            \draw[->]   (X1) edge (X2);
        \end{tikzpicture}
    \caption{Known PAG used in Example \ref{ex:combine-limits}}
    \label{fig:ex-combine-limits}
\end{figure}

\begin{example}[Limitations of R1 Combine]
\label{ex:combine-limits}
    Consider an unknown causal model over $\{V_1, V_2, V_3, X_1, X_2, Y\}$, where we have data on every variable and expert knowledge that ${\{V_1, V_2, V_3\} < \{X_1,X_2\} < Y}$. Let the PAG in Figure \ref{fig:ex-combine-limits} represent all the in/dependencies we can learn from the data with the addition of our expert knowledge. We want to know the effect of $\{X_1,X_2\}$ on $Y$. 
    
    Using graphical criteria from prior research (see Theorem \ref{thm:as-graphical-pags} in Supp. \ref{supp:prelims}), we can show there is an adjustment set relative to $(\{X_1, X_2\}, Y)$ -- for example, $\{V_1,V_2\}$. But Theorem \ref{thm:r1-combine} cannot identify any such set, since there is no $W_1 \in \{V_1, V_2, V_3\}$ and $\mathbf{T_1} \subseteq \{V_1, V_2, V_3\}$ such that $(W_1 \perp Y | \mathbf{T_1}, X_1)$.
\end{example}

% --------------  LIMITATION 2  ---------------- %

A second limitation of our methods is computational. Theorems \ref{thm:r1-build} and \ref{thm:r1-combine} require a search for variables $W_i$ that fulfill specific in/dependencies, and further, Theorem \ref{thm:r1-combine} requires additional in/dependence testing in a search for minimal adjustment sets (see Lemma \ref{lem:minimal-criteria}). To do this in our simulations, we performed brute-force searches over all possible variables. But these searches, which are expensive and require a large number of hypothesis tests, may hinder the applicability of our methods in high-dimensional settings.

Certainly, future work could address this by exploring greedy-search approaches for implementation. But we want to highlight an alternative -- that an expensive brute-force search may not be necessary in practice, since expert domain knowledge could guide a search for an adjustment set that fits the criteria of Theorems 5 and 7. That is, researchers could begin such a search based on their experiences in the field, and thus, may not need to test all possible combinations of nodes before selecting an adjustment set that identifies their causal effect of interest. For a clear example of this, consider when an adjustment set for one treatment is already known and researchers are interested in modifying their analysis to account for additional treatments. Our methods are well-suited for this setting.

% --------------  LIMITATION 3  ---------------- %

A third limitation of our work is that our methods cannot conclude that there is no treatment effect. Thus, we saw in our simulation study (Section \ref{sec:simulations}) that an existing approach outperformed our methods in Setting 2, where datasets are generated from models without a treatment effect. EHS \cite{entner2013data} address this setting with a second rule (R2) that can obtain a \textit{negative} outcome. Future research could extend R2 to a setting with multiple treatments, and combining such an extension with our extensions of R1 could improve estimation of a ``causal effect'' of zero.

In addition to addressing the above limitations, future work could consider  extensions to conditional adjustment \citep{laplante2023conditional}, or combining experimental and observational datasets for estimation \citep{triantafillou2021learning, triantafillou2021causal, triantafillou2023learning}. Further extensions that allow for inclusions of mediating variables may also be possible, for instance by considering sequential (time-dependent) back-door adjustment sets \cite{pearl1995probabilistic, murphy2003optimal, pearl2009causality, smucler2022efficient}.

%%%%%%%%%%%%%%%%%%%%%%%%%%%%%%%%%%%%
%         JCI STATEMENTS           %
%%%%%%%%%%%%%%%%%%%%%%%%%%%%%%%%%%%%

%\vskip .2in
%\noindent
%\textbf{Acknowledgments:}
%The authors...

\vskip .2in
\noindent
\textbf{Funding information:}
This material is based upon work supported by the National Science Foundation under Grant No. 2210210.

\vskip .2in
\noindent
\textbf{Author contributions:}
All authors have accepted responsibility for the entire content of this manuscript and consented to its submission to the journal, reviewed all the results, and approved the final version of the manuscript. EP conceived the project, and all authors collaborated to develop the methods. SL designed and wrote all proofs, conducted the literature review, and drafted the manuscript. ST designed and implemented the simulation study. All authors discussed and reviewed each other's work.

\vskip .2in
\noindent
\textbf{Conflict of interest:}
Authors state no conflict of interest.

\vskip .2in
\noindent
\textbf{Data availability statement:}
The code used to generate and analyze datasets for our simulation study is available at https://github.com/striantafillou.

%----------------

\bibliography{main}

%%%%%%%%%%%%%%%%%%%%%%%%%%%%%%%%%%%%%%%%%%%%%%%%%%%
%%%%%%%%%%%%%%%%%%%%%%%%%%%%%%%%%%%%%%%%%%%%%%%%%%%
%%%%%%%%%%%%%%%%%%%%%%%%%%%%%%%%%%%%%%%%%%%%%%%%%%%
%%%%%%%%%%%%%%%%%%%%%%%%%%%%%%%%%%%%%%%%%%%%%%%%%%%
%%%%%%%%%%%%%%%%%%%%%%%%%%%%%%%%%%%%%%%%%%%%%%%%%%%
%%%%%%%%%%%%%%%%%%%%%%%%%%%%%%%%%%%%%%%%%%%%%%%%%%%
%%%%%%%%%%%%%%%%%%%%%%%%%%%%%%%%%%%%%%%%%%%%%%%%%%%

% ======================================================
%
%                 SUPPLEMENTARY MATERIAL
%
% ====================================================== 

\clearpage
\appendix

\centerline{\LARGE{Supplement to:}}
\centerline{\LARGE{Data-Driven Adjustment for Multiple Treatments}}

\tableofcontents

%%%%%%%%%%%%%%%%%%%%%%%%%%%%%%%%%%%%
%           SUPPLEMENT A             %
%%%%%%%%%%%%%%%%%%%%%%%%%%%%%%%%%%%%

\section{Further Preliminaries}
\label{supp:prelims}

% -------------  DAGS  -------------%
\subsection{Directed Graphs}

\textbf{Proper and Back-door Paths.}
A path from $\mb{X}$ to $\mb{Y}$ is \textit{proper} (with respect to $\mb{X}$) if only its first node is in $\mb{X}$. A path from $X$ to $Y$ that begins with the edge $X \gets$ is said to be a path \textit{into} $X$, or a \textit{back-door path}.

\begin{definition}
\label{def:bas-graphical}
{\normalfont (\textbf{Back-door Adjustment Set for DAGs}; \cite{maathuis2015generalized})}
    Let $\mb{X}$, $\mb{Y}$, and $\mb{Z}$ be pairwise disjoint node sets in a DAG $\g[D]$. Then $\mb{Z}$ is a back-door adjustment set relative to $(\mb{X,Y})$ in $\g[D]$ if and only if:
    \begin{enumerate}[label=\emph{(\alph*)}]
        \item $\mb{Z} \cap \De(\mb{X},\g[D]) = \emptyset$, and 
        \item $\mb{Z} \cup \big[ \mb{X} \setminus \{X\} \big]$ blocks all back-door paths from $X$ to $\mb{Y}$ in $\g[D]$, for all $X \in \mb{X}$.
    \end{enumerate}
\end{definition}

\begin{theorem}
\label{thm:as-graphical}
{\normalfont (\textbf{Adjustment Set for DAGs, Graphical Criteria}; \citep{shpitser2012validity, perkovic2018complete})}
    Let $\mb{X}$, $\mb{Y}$, and $\mb{Z}$ be pairwise disjoint node sets in a causal DAG $\g[D]$. Then $\mb{Z}$ is an adjustment set relative to $(\mb{X,Y})$ in $\g[D]$ if and only if:
    \begin{enumerate}[label=\emph{(\alph*)}]
        \item $\mb{Z}$ contains no descendants of any $W \notin \mb{X}$ that lies on a proper causal path from $\mb{X}$ to $\mb{Y}$ in $\g[D]$, and 
        \item $\mb{Z}$ blocks all proper non-causal paths from $\mb{X}$ to $\mb{Y}$ in $\g[D]$.
    \end{enumerate}
\end{theorem}

\begin{lemma}
\label{lem:as-bas} 
{\normalfont (cf. Theorem 3.1 of \cite{maathuis2015generalized})}
    Let $\mb{X}$, $\mb{Y}$, and $\mb{Z}$ be pairwise disjoint node sets in a causal DAG $\g[D]$. If $\mb{Z}$ is a back-door adjustment set relative to $(\mb{X,Y})$ in $\g[D]$, then $\mb{Z}$ is an adjustment set relative to $(\mb{X,Y})$ in $\g[D]$.
\end{lemma}

% -------------  PAGS  -------------%
\subsection{Ancestral Graphs}

The following are key definitions related to ancestral graphs and their associated densities. We rely on the framework of \cite{richardson2002ancestral, zhang2008completeness, ali2009markov}.

\textbf{Mixed and Partially Directed Mixed Graphs.}
A \textit{mixed graph} may contain directed ($\to$) and bi-directed ($\leftrightarrow$) edges. The \textit{partially directed mixed graphs} we consider may contain directed, bi-directed, undirected ($\circcirc$), or partially directed ($\circarrow$) edges. We use $\bullet$ as a stand in for any edge mark.

\textbf{Definite Status Paths.}
Let $\g$ be a mixed or partially directed mixed graph with a path $p:=\langle X_1, \dots, X_k \rangle$, $k>1$. If $p$ contains $X_{j-1} \bulletarrow X_j \arrowbullet X_{j+1}$ for $1<j<k$, then $X_j$ is a \textit{collider} on $p$. $X_j$ is a \textit{definite non-collider} on $p$ if $p$ contains $X_{j-1} \leftarrow X_j$ or $X_j \rightarrow X_{j+1}$, or if $\g$ contains $X_{j-1} \circcirc X_j \circcirc X_{j+1}$ but no edge $\langle X_{j-1}, X_{j+1} \rangle$. If every node on $p$ is a collider, definite non-collider, or endpoint on $p$, then $p$ is a \textit{definite status path}. 

\textbf{M-connection and M-separation.}
Let $\mb{X}$, $\mb{Y}$, and $\mb{Z}$ be pairwise disjoint node sets in a mixed or partially directed mixed graph $\g$. A definite-status path $p$ from $\mb{X}$ to $\mb{Y}$ in $\g$ is \textit{open} given $\mb{Z}$ if every definite non-collider on $p$ is not in $\mb{Z}$ and every collider on $p$ has a descendant in $\mb{Z}$ in $\g$. Otherwise, $p$ is \textit{blocked} given $\mb{Z}$. If $\mb{Z}$ blocks all definite-status paths between $\mb{X}$ and $\mb{Y}$ in $\g$, then $\mb{X}$ is \textit{m-separated} from $\mb{Y}$ given $\mb{Z}$ in $\g$ and we write ${(\mb{X} \msepp \mb{Y} \,|\, \mb{Z})_{\g}}$. Otherwise, $\mb{X}$ is \textit{m-connected} to $\mb{Y}$ given $\mb{Z}$ in $\g$ and we write ${(\mb{X} \not\msepp \mb{Y} \,|\, \mb{Z})_{\g}}$.

\textbf{MAGs.}
A directed path from $X$ to $Y$ and the edge $X \to Y$ form an \textit{almost directed cycle}. A mixed graph without directed or almost directed cycles is called \textit{ancestral}. Note that we do not consider ancestral graphs that represent selection bias. A \textit{maximal ancestral graph} (MAG) is an ancestral graph $\g[M] = (\mb{V,E})$ where every pair of non-adjacent nodes $X$ and $Y$ in $\g[M]$ can be m-separated by a set $\mb{Z} \subseteq \mb{V} \setminus \{X,Y\}$. A DAG $\g[D] = (\mb{V,E})$ with unobserved variables $\mb{U} \subseteq \mb{V}$ can be uniquely \textit{represented by} a MAG $\g[M] = (\mb{V} \setminus \mb{U}, \mb{E'})$, which preserves the ancestry and m-separations among the observed variables.

\textbf{PAGs.}
All MAGs that encode the same set of m-separations form a Markov equivalence class, which can be uniquely \textit{represented by} a partially directed mixed graph called a \textit{partial ancestral graph} (PAG). $[\g]$ denotes all MAGs represented by a PAG $\g$. We say a DAG $\g[D]$ is \textit{represented by} a PAG $\g$ if there is a MAG $\g[M] \in [\g]$ such that $\g[D]$ is represented by $\g[M]$. Note that we only consider maximally informative PAGs that are complete with respect to orientation rules $R1-R4$ and $R8-R10$ of Zhang \citep{zhang2008completeness} and that do not represent selection bias.

%================================

\textbf{Markov Compatibility and Faithfulness.}
We say an observational density is \textit{Markov compatible} with a MAG or PAG $\g$ if it is Markov compatible with a DAG represented by $\g$. We say an observational density is \textit{faithful} to a MAG or PAG $\g$ if it is faithful to a DAG represented by $\g$.

\textbf{Probabilistic Implications of a Graph.}
Let $\mb{X}$, $\mb{Y}$, and $\mb{Z}$ be pairwise disjoint node sets in a MAG or PAG $\g$. If ${(\mb{X} \msepp \mb{Y} \,|\, \mb{Z})_{\g}}$, then $\mb{X}$ and $\mb{Y}$ are independent given $\mb{Z}$ in any observational density that is Markov compatible with $\g$. If ${(\mb{X} \not\msepp \mb{Y} \,|\, \mb{Z})_{\g}}$, then $\mb{X}$ and $\mb{Y}$ are dependent given $\mb{Z}$ in any observational density that is faithful to $\g$.

\textbf{Causal Graphs.}
Let $\g$ be a graph with nodes $V_i$ and $V_j$. When $\g$ is a MAG or PAG, it is a \textit{causal MAG} or \textit{causal PAG} if every edge $V_i \to V_j$ represents the presence of a causal path from $V_i$ to $V_j$; every edge $V_i \arrowbullet V_j$ represents the absence of a causal path from $V_i$ to $V_j$; and every edge $V_i \circcirc V_j$ represents the presence of a causal path of unknown direction or a common cause in the underlying causal DAG. 

\textbf{Possibly Causal Paths.}
Let $p:=\langle X_1, \dots, X_k \rangle$, $k>1$, be a path in a causal MAG or PAG $\g$. If $\g$ does not contain an edge $X_i \arrowbullet X_j, 1 \le i < j \le k$, then $p$ is \textit{possibly causal} and $X_2, \dots, X_k$ are \textit{possible descendants} of $X_1$. Otherwise, $p$ is \textit{non-causal}.

%================================

\textbf{Visible Edges.}
Let $\g$ be a MAG or PAG. We denote that $X$ is adjacent to $Y$ in $\g$ by $X \in \Adj(Y,\g)$. A directed edge $X \rightarrow Y$ is \textit{visible} in $\g$ if there is a node $V \notin \Adj(Y, \g)$ such that $\g$ contains either $V \bulletarrow X$ or $V \bulletarrow V_1 \leftrightarrow \dots \leftrightarrow V_k \leftrightarrow X$, where $k \ge 1$ and $V_1, \dots, V_k \in \Pa(Y, \g) \setminus \{V,X,Y\}$.

\textbf{Consistency.}
We say an interventional density is \textit{consistent} with a causal MAG or PAG $\g$ if it is consistent with each DAG represented by $\g$ -- were the DAG to be causal.

\textbf{Adjustment Sets.}
Let $\mb{X}$, $\mb{Y}$, and $\mb{Z}$ be pairwise disjoint node sets in a causal MAG or PAG $\g$. Then $\mb{Z}$ is an adjustment set relative to $(\mb{X,Y})$ in $\g$ if and only if $f(\mb{y}|do(\mb{x})) = \int f(\mb{y} | \mb{x}, \mb{z}) f(\mb{z}) \diff \mb{z}$ for any $f$ consistent with $\g$. We omit reference to $(\mb{X,Y})$ or $\g$ when it can be assumed.

% -------------  EXISTING RESULTS  -------------%

\begin{theorem}
\label{thm:as-graphical-pags}
{\normalfont (\textbf{Adjustment Set for PAGs, Graphical Criteria}; cf. Theorem 5 of \cite{perkovic2018complete})}
    Let $\mb{X}$, $\mb{Y}$, and $\mb{Z}$ be pairwise disjoint node sets in a causal PAG $\g$. Then $\mb{Z}$ is an adjustment set relative to $(\mb{X},\mb{Y})$ in $\g$ if and only if
    \begin{enumerate}[label=\emph{(\alph*)}]
        \item every proper, possibly causal path from $\mb{X}$ to $\mb{Y}$ in $\g$ starts with a visible edge,
        \item $\mb{Z}$ contains no possible descendants of any $W \notin \mb{X}$ that lies on a proper, possibly causal path from $\mb{X}$ to $\mb{Y}$ in $\g$, and 
        \item $\mb{Z}$ blocks all proper, definite status, non-causal paths from $\mb{X}$ to $\mb{Y}$ in $\g$.
    \end{enumerate}
\end{theorem}

%%%%%%%%%%%%%%%%%%%%%%%%%%%%%%%%%%%%
%           SUPPLEMENT B             %
%%%%%%%%%%%%%%%%%%%%%%%%%%%%%%%%%%%%

\section{Proof for Section \ref{sec:r1-build}: R1 Build}
\label{supp:r1-build}

% -------------  MAIN PROOF  -------------%

\begin{proofofnoqed}[Theorem \ref{thm:r1-build} (R1 Build)]
    The result holds for $k=1$ by Theorem \ref{thm:r1-entner}. Thus, let $k \ge 2$. Suppose there exist $W_1, \dots, W_k \in \mb{W}$ and $\mb{Z} \subseteq \mb{W} \setminus \{W_1, \dots, W_k\}$ such that \ref{thm:r1-build-i} and \ref{thm:r1-build-ii} hold for $i \in \{1, \dots, k\}$. Let $\g[D]$ be the DAG induced by the causal model. Then for ease of notation, let $\mb{X^N_i} = \mb{X} \setminus \De(X_i, \g[D])$, and note that $\{X_1, \dots, X_i\}^{-i} \subseteq \mb{X^N_i}$ since $X_1 < \dots < X_k$ is a causal ordering consistent with the model.

    We start by showing that $\mb{X}$ has a causal effect on $Y$. Consider an arbitrary $j \in \{1,\dots, k\}$. By \ref{thm:r1-build-i}, \ref{thm:r1-build-ii}, and faithfulness, there must be a path $p_j$ from $W_j$ to $Y$ in $\g[D]$ that is open given $\mb{Z} \cup \{X_1, \dots, X_j\}^{-j}$ and contains $X_j$ as a non-collider. Therefore, either $p_j(W_j,X_j)$ ends $\gets X_j$ or $p_j(X_j,Y)$ begins $X_j \to$. For sake of contradiction, suppose the former holds. Since $X_j \notin \An(\mb{W}, \g[D])$, then $p_j(W_j,X_j)$ must contain a collider. But since $p_j$ is open given $\mb{Z} \cup \{X_1, \dots, X_j\}^{-j} \subseteq \mb{W} \cup \mb{X_j^N}$, the closest collider to $X_j$ on $p_j(W_j,X_j)$ must be in $\An( \mb{Z} \cup \{X_1, \dots, X_j\}^{-j}, \,\,\g[D] )$, which contradicts that $X_j \notin \An(\mb{W} \cup \mb{X_j^N}, \g[D])$. Thus, $p_j(X_j,Y)$ must begin $X_j \to$. By similar logic, $p_j(X_j,Y)$ cannot contain a collider, and so $p_j$ must end $\to X_j \to \dots \to Y$.

    We continue by showing that $\mb{Z}$ satisfies the conditions of Definition \ref{def:bas-graphical} and thus by Lemma \ref{lem:as-bas}, is an adjustment set relative to $(\mb{X},Y)$. By assumption, no variable in $\mb{Z}$ is a descendant of $\mb{X}$. Thus, we only need to show the following for $i \in \{1, \dots, k\}$:
    \begin{align*}
        \mb{Z} \cup \mb{X^{\text{-}i}} \text{ blocks ever}&\text{y back-door path} \label{thm:r1-build-star} \tag{$\ast$}\\
        \text{from } X_i \text{ to}&\text{ } Y \text{ in } \g[D]. 
    \end{align*}
    We begin with $i=k$, and proceed with a proof by induction for $i \in \{1, \dots, k-1\}$.

    %----- BDP X2 to Y -----%

    \textbf{BASE CASE: }
    Note from the discussion above that there must be a path $p_k$ from $W_k$ to $Y$ in $\g[D]$ that is open given $\mb{Z} \cup \mb{X}^{-k}$ and ends $\to X_k \to \dots \to Y$. Then for sake of contradiction, suppose there is a back-door path $q_k$ from $X_k$ to $Y$ in $\g[D]$ that is open given $\mb{Z} \cup \mb{X^{\text{-}k}}$. Let $A$ be the node on both $p_k$ and $q_k$ that is closest to $W_k$ on $p_k$, and define $r_k = p_k(W_k, A) \oplus q_k(A,Y)$. We show below that $r_k$ must be open given $\mb{Z} \cup \mb{X}$ -- that is, no non-collider on $r_k$ is in $\mb{Z} \cup \mb{X}$ and every collider on $r_k$ is in $\An\big(\mb{Z} \cup \mb{X}, \g[D] \big)$ -- which contradicts \ref{thm:r1-build-ii} by faithfulness. 

    First consider the nodes on $r_k$ other than $A$. Every collider on $r_k(W_k,A)$ and $r_k(A, Y)$ is in $\An\big( \mb{Z} \cup \mb{X}, \g[D] \big)$, since $p_k$ and $q_k$ are open given $\mb{Z} \cup \mb{X^{\text{-}k}}$. By the same logic, no non-collider on $r_k(W_k,A)$ or $r_k(A, Y)$ is in $\mb{Z} \cup \mb{X^{\text{-}k}}$. Further, $X_k$ is not a non-collider on $r_k(W_k,A)$ or $r_k(A, Y)$, since $r_k$ does not contain $X_k$ except possibly $A=X_k$.

    Next consider the node $A$. When $A = W_k$, note that $r_k = r_k(A,Y)$, and the base case is done. When $A \neq W_k$, we show in the cases below that $A$ is either a collider on $r_k$ such that $A \in \An\big(\mb{Z} \cup \mb{X}, \g[D]\big)$ or a non-collider on $r_k$ such that $A \notin \mb{Z} \cup \mb{X}$, which completes the base case.
    \begin{itemize}
        \item
        Let $A=X_k$. Note that $A$ is a collider on $r_k$ (by definition of $p_k$ and $q_k$) and $A \in \An\big(\mb{Z} \cup \mb{X}, \g[D]\big)$.

        \item
        Let $A \in \mb{Z} \cup \mb{X^{\text{-}k}}$. Since $p_k$ and $q_k$ are open given $\mb{Z} \cup \mb{X^{\text{-}k}}$, then $A$ must be a collider on both paths. Therefore, $A$ is a collider on $r_k$ and $A \in \An\big(\mb{Z} \cup \mb{X}, \g[D]\big)$.

        \item
        Let $A \notin \mb{Z} \cup \mb{X}$. When $A$ is a non-collider on $r_k$, the claim holds. When $A$ is a collider on $r_k$, we only need to show that $A \in \An( \mb{Z} \cup \mb{X}, \g[D] )$. This holds if $A$ is a collider on $p_k$, since $p_k$ is open given $\mb{Z} \cup \mb{X^{\text{-}k}}$. Consider when $A$ is a collider on $r_k$ and a non-collider on $p_k$. Note that $p_k(W_k, X_k)$ begins $W_k \dots \to A \to$ and ends $\to X_k$. When $p_k(A, X_k)$ is directed, then $A \in \An( \mb{Z} \cup \mb{X}, \g[D] )$. When $p_k(A, X_k)$ contains a collider, the earliest such collider must be in $\An( \mb{Z} \cup \mb{X^{\text{-}k}} , \g[D] )$, since $p_k$ is open given $\mb{Z} \cup \mb{X^{\text{-}k}}$. Thus, $A \in \An( \mb{Z} \cup \mb{X}, \g[D] )$.
    \end{itemize}

    %----- BDP X1 to Y -----%

    \textbf{INDUCTION:}
    Pick an arbitrary $j \in \{1, \dots, k-1\}$, and for sake of induction, assume that \eqref{thm:r1-build-star} holds for $i \in \{j+1, \dots, k\}$. We will show that \eqref{thm:r1-build-star} also holds for $i=j$.

    Recall that there must be a path $p_j$ from $W_j$ to $Y$ in $\g[D]$ that is open given $\mb{Z} \cup \{X_1, \dots, X_j\}^{-j}$ and ends $\to X_j \to \dots \to Y$. Then for sake of contradiction, suppose there is a back-door path $q_j$ from $X_j$ to $Y$ in $\g[D]$ that is open given $\mb{Z} \cup \mb{X^{\text{-}j}}$. Let $B$ be the node on both $p_j$ and $q_j$ that is closest to $W_j$ on $p_j$, and define $r_j = p_j(W_j, B) \oplus q_j(B,Y)$. We show below that $r_j$ must be open given $\mb{Z} \cup \{X_1, \dots, X_j\}$ -- that is, no non-collider on $r_j$ is in $\mb{Z} \cup \{X_1, \dots, X_j\}$ and every collider on $r_j$ is in $\An\big( \mb{Z} \cup \{X_1, \dots, X_j\}, \g[D] \big)$ -- which contradicts \ref{thm:r1-build-ii} by faithfulness.
	
    First consider the nodes on $r_j(W_j,B)$. Every collider on $r_j(W_j,B)$ is in $\An\big( \mb{Z} \cup \{X_1, \dots, X_j\}, \g[D] \big)$, since $p_j$ is open given $\mb{Z} \cup \{X_1, \dots, X_{j}\}^{\text{-}j}$. By the same logic and the fact that $r_j$ does not contain $X_j$ except possibly $B=X_j$, no non-collider on $r_j(W_j,B)$ is in $\mb{Z} \cup \{X_1, \dots, X_{j}\}$.

    Next consider the nodes on $r_j(B,Y)$. No non-collider on $r_j(B,Y)$ is in $\mb{Z} \cup \{X_1, \dots, X_j\}$, since $q_j$ is open given $\mb{Z} \cup \mb{X^{\text{-}j}}$ and $X_j$ is an endpoint on $q_j$. Then suppose for sake of contradiction that there is a collider on $r_j(B,Y)$ not in $\An\big( \mb{Z} \cup \{X_1, \dots, X_j\}, \g[D] \big)$, and let $C$ be the closest such collider to $Y$ on $r_j(B,Y)$. Note that $C \in \An( \{X_{j+1}, \dots, X_k\}, \g[D] )$, since $q_j$ is open given $\mb{Z} \cup \mb{X^{\text{-}j}}$. Thus, let $s$ be a shortest directed path in $\g[D]$ from $C$ to $\{X_{j+1}, \dots, X_k\}$, where we denote the latter endpoint $X_\ell$, $\ell \in \{j+1, \dots, k\}$. Then let $E$ be the node on both $s$ and $r_j$ that is closest to $Y$ on $r_j$.

    Consider the path $t = s(X_\ell,E) \oplus r_j(E,Y)$. Start by noting that no non-collider on $t$ is in $\mb{Z} \cup \mb{X^{\text{-}\ell}}$ due to the following: $C \notin \An\big( \mb{Z} \cup \{X_1, \dots, X_j\}, \g[D] \big)$; $s$ is a shortest path to $\{X_{j+1}, \dots, X_k\}$; $q_j$ is open given $\mb{Z} \cup \mb{X^{\text{-}j}}$; and $X_j$ is an endpoint on $q_j$. Further, every collider on $t$ must be in $\An( \mb{Z} \cup \mb{X^{\text{-}\ell}}, \g[D] )$, since $t$ can only contain colliders that are colliders on $r_j(E,Y)$, where by definition of $C$, every collider on $r_j(E,Y)$ is in $\An\big( \mb{Z} \cup \{X_1, \dots, X_j\}, \g[D] \big)$. Thus, $t$ -- a back-door path from $X_\ell$ to $Y$ -- is open given $\mb{Z} \cup \mb{X^{\text{-}\ell}}$, which contradicts our induction assumption that \eqref{thm:r1-build-star} holds for $i=\ell$, $\ell \in \{j+1, \dots, k\}$.

    Finally, consider the node $B$. When $B = W_j$, note that $r_j = r_j(B,Y)$, and the induction step is done. When $B \neq W_j$, we show in the cases below that $B$ is either a collider on $r_j$ such that $B \in \An\big(\mb{Z} \cup \{X_1, \dots, X_j\}, \g[D]\big)$ or a non-collider on $r_j$ such that $B \notin \mb{Z} \cup \{X_1, \dots, X_j\}$, which completes the proof.
    \begin{itemize}
        \item
        Let $B=X_j$. Note that $B$ is a collider on $r_j$ (by definition of $p_j$ and $q_j$) and $B \in \An\big(\mb{Z} \cup \{X_1, \dots, X_j\}, \g[D]\big)$.

        \item
        Let $B \in \mb{Z} \cup \{X_1, \dots, X_{j}\}^{\text{-}j}$. Since $p_j$ is open given $\mb{Z} \cup \{X_1, \dots, X_{j}\}^{\text{-}j}$ and $q_j$ is open given $\mb{Z} \cup \mb{X^{\text{-}j}}$, then $B$ must be a collider on both paths. Therefore, $B$ is a collider on $r_j$ and $B \in \An\big(\mb{Z} \cup \{X_1, \dots, X_j\}, \g[D]\big)$.

        \item
        For sake of contradiction, let $B = X_\ell$, $\ell \in \{j+1, \dots, k\}$. Note that $B$ must be a collider on $q_j$, since $q_j$ is open given $\mb{Z} \cup \mb{X^{\text{-}j}}$. Thus, $r_j(B,Y)$ is a back-door path from $X_\ell$ to $Y$. However, note that no non-collider on $r_j(B,Y)$ is in $\mb{Z} \cup \mb{X^{\text{-}\ell}}$, since $q_j$ is open given $\mb{Z} \cup \mb{X^{\text{-}j}}$ and since $X_j$ is an endpoint on $q_j$. Further, every collider on $r_j(B,Y)$ is in $\An\big( \mb{Z} \cup \mb{X^{\text{-}\ell}}, \g[D] \big)$, since we have shown that every collider on $r_j(B,Y)$ is in $\An\big( \mb{Z} \cup \{X_1, \dots, X_j\}, \g[D] \big)$. Therefore, $r_j(B,Y)$ is a back-door path from $X_\ell$ to $Y$ that is open given $\mb{Z} \cup \mb{X^{\text{-}\ell}}$, which contradicts our induction assumption that \eqref{thm:r1-build-star} holds for $i=\ell$, $\ell \in \{j+1, \dots, k\}$.

        \item
        Let $B \notin \mb{Z} \cup \mb{X}$. When $B$ is a non-collider on $r_j$, the claim holds. When $B$ is a collider on $r_j$, we only need to show that $B \in \An\big( \mb{Z} \cup \{X_1, \dots, X_j\}, \g[D] \big)$. This holds if $B$ is also a collider on $p_j$, since $p_j$ is open given $\mb{Z} \cup \{X_1, \dots, X_{j}\}^{\text{-}j}$. Consider when $B$ is a collider on $r_j$ and a non-collider on $p_j$. Note that $p_j(W_j, X_j)$ begins $W_j \dots \to B \to$ and ends $\to X_j$. When $p_j(B, X_j)$ is directed, then $B \in \An\big( \mb{Z} \cup \{X_1, \dots, X_j\}, \g[D] \big)$. When $p_j(B, X_j)$ contains a collider, the earliest such collider must be in $\An( \mb{Z} \cup \{X_1, \dots, X_{j}\}^{\text{-}j} , \g[D] )$, since $p_j$ is open given $\mb{Z} \cup \{X_1, \dots, X_{j}\}^{\text{-}j}$. Thus, $B \in \An\big( \mb{Z} \cup \{X_1, \dots, X_j\}, \g[D] \big)$. \hfill\BlackBox
    \end{itemize}
\end{proofofnoqed}

%%%%%%%%%%%%%%%%%%%%%%%%%%%%%%%%%%%%
%           SUPPLEMENT C             %
%%%%%%%%%%%%%%%%%%%%%%%%%%%%%%%%%%%%

\section{Proofs for Section \ref{sec:r1-combine}: R1 Combine}
\label{supp:r1-combine}

% -------------  MAIN RESULTS  -------------%
\subsection{Main Results}

% --- R1 COMBINE ---%
\begin{proofof}[Theorem \ref{thm:r1-combine} (R1 Combine)]
    The result holds for $k=1$ by Theorem \ref{thm:r1-entner}. Thus, let $k \ge 2$. Suppose there exist $W_1, \dots, W_k \in \mb{W}$ and $\mb{T_i} \subseteq \big[ \mb{W} \setminus \{W_i \} \big] \cup \mb{X^N_i}$ such that \ref{thm:r1-combine-i}-\ref{thm:r1-combine-ii} hold for all $i \in \{1,\dots, k\}$. Note by \ref{thm:r1-combine-i}, \ref{thm:r1-combine-ii}, and Theorem \ref{thm:r1-entner} that $X_i$ has a causal effect on $Y$ that is identifiable through the adjustment set $\mb{T_i}$. Thus, $\mb{X}$ has a causal effect on $Y$, and we can define $\mb{Z_i}$ to be any minimal adjustment set relative to $(X_i,Y)$ such that $\mb{Z_i} \subseteq \mb{T_i}$. Let $\g[D]$ be the DAG induced by the causal model so that $\mb{X^N_i} = \mb{X} \setminus \De(X_i, \g[D])$. Then for ease of notation, let $\mb{X^{\text{-}i}} = \mb{X} \setminus \{X_i\}$.

    We show below that $\mb{Z} := \cup_{i=1}^k \mb{Z_i} \setminus \mb{X}$ satisfies the conditions of Definition \ref{def:bas-graphical} and thus by Lemma \ref{lem:as-bas}, is an adjustment set relative to $(\mb{X},Y)$. Since $\mb{Z} \subseteq \mb{W}$ where $\mb{W} < \mb{X}$, we only need to show for every $i \in \{1, \dots, k\}$ that $\mb{Z} \cup \mb{X^{\text{-}i}}$ blocks every back-door path from $X_i$ to $Y$ in $\g[D]$. Without loss of generality, we show this holds for $i=1$.

    For sake of contradiction, suppose there is a back-door path $q$ from $X_1$ to $Y$ in $\g[D]$ that is open given $\mb{Z} \cup \mb{X^{\text{-}1}}$. Since $\mb{Z_1}$ is an adjustment set relative to $(X_1,Y)$, then by Theorem \ref{thm:as-graphical}, $q$ must be blocked given $\mb{Z_1} \subseteq \mb{Z} \cup \mb{X^{\text{-}1}}$. This implies that $q$ must contain a collider in $\An\big(\mb{Z} \cup \mb{X^{\text{-}1}}, \g[D] \big) \setminus \An(\mb{Z_1}, \g[D])$. Let $C_1, \dots, C_\ell$, $\ell \ge 1$, be the set of all such colliders -- ordered so that $C_1$ is the closest such collider to $X_1$ on $q$ and $C_\ell$ is the furthest such collider from $X_1$ on $q$.

    We pause to define a path $t_i$ in $\g[D]$ that is directed from $C_i$ to $Y$ for any $i \in \{1, \dots, \ell\}$ such that $C_i \notin \An(X_1, \g[D])$. This path will be useful in showing a contradiction. First pick such an $i \in \{1, \dots, \ell\}$ (supposing one exists), and define $r_i$ as a longest directed path from $C_i$ to $\big(\mb{Z} \cup \mb{X^{\text{-}1}} \big) \setminus \An(\mb{Z_1}, \g[D])$ in $\g[D]$. Then define $t_i$ based on how $r_i$ ends. 

    When $r_i$ ends with $X_j \in \mb{X^{\text{-}1}} \setminus \An(\mb{Z_1},\g[D])$ for some $j \in \{2, \dots, k\}$, note that by \ref{thm:r1-combine-i}, \ref{thm:r1-combine-ii}, and faithfulness, there must be a path $p_j$ from $W_j$ to $Y$ in $\g[D]$ that is open given $\mb{T_j}$ and contains $X_j$ as a non-collider. Therefore, either $p_j(W_j,X_j)$ ends $\gets X_j$ or $p_j(X_j,Y)$ begins $X_j \to$. For sake of contradiction, suppose the former holds. Since $X_j \notin \An(\mb{W}, \g[D])$, then $p_j(W_j,X_j)$ must contain a collider. But since $p_j$ is open given $\mb{T_j} \subseteq \mb{W} \cup \mb{X_j^N}$, the closest collider to $X_j$ on $p_j(W_j,X_j)$ must be in $\An(\mb{T_j}, \g[D])$, which contradicts that $X_j \notin \An(\mb{W} \cup \mb{X_j^N}, \g[D])$. Thus, $p_j(X_j,Y)$ must begin $X_j \to$. By similar logic, $p_j(X_j,Y)$ cannot contain a collider, and so $p_j(X_j,Y)$ must take the form $X_j \to \dots \to Y$. Thus, we can define $t_i = r_i \oplus p_j(X_j, Y)$.

    Next, consider when $r_i$ ends with $Z_j \in \mb{Z} \setminus \An(\mb{Z_1}, \g[D])$ so that $Z_j \in \mb{Z_j} \setminus \mb{X}$ for some $j \in \{2, \dots, k\}$. By definition, $\mb{Z_j}$ is an adjustment set relative to $(X_j,Y)$, but this does not hold for any subset $\mb{Z_j} \setminus \{Z\}$, where $Z \in \mb{Z_j}$. Thus by Theorem \ref{thm:as-graphical}, there must be a non-causal path from $X_j$ to $Y$ in $\g[D]$ that contains $Z_j$ as a non-collider and that is open given $\mb{Z_j} \setminus \{Z_j\}$. Let $s_j$ be one such path. For sake of contradiction, suppose $s_j(Z_j,Y)$ begins $Z_j \gets$. Since $Z_j$ is a non-collider on $s_j$, then $s_j(X_j, Z_j)$ must end with $\gets Z_j$. If $s_j(X_j, Z_j)$ contains a collider, then by the definition of $s_j$, the closest such collider to $Z_j$ on $s_j$ must be in $\An( \mb{Z_j} \setminus \{Z_j\}, \g[D])$, which contradicts either the definition of $r_i$ as a longest path to $\big(\mb{Z} \cup \mb{X^{\text{-}1}} \big) \setminus \An(\mb{Z_1}, \g[D])$ or the definition of $C_i \notin \An(\mb{Z_1} \cup \{X_1\},\g[D])$. The same contradiction holds if instead $s_j(X_j, Z_j)$ takes the form $X_j \gets \dots \gets Z_j$. Therefore, $s_j(Z_j, Y)$ must begin $Z_j \to$. Further, $s_j(Z_j,Y)$ must take the form $Z_j \to \dots \to Y$, since by the same logic, $s_j(Z_j,Y)$ cannot contain a collider. Thus, here we define $t_i = r_i \oplus s_j(Z_j,Y)$.

    We proceed to cases with a directed path from $C_i$ to $Y$ defined as follows:
    \begin{align*}
        t_i := 
        \begin{cases}
            r_i \oplus p_j(X_j, Y)   &r_i \text{ ends } X_j \in \mb{X^{\text{-}1}} \setminus \An(\mb{Z_1},\g[D]) \\
            r_i \oplus s_j(Z_j,Y)    &r_i \text{ ends } Z_j \in \mb{Z} \setminus \An(\mb{Z_1}, \g[D]).
        \end{cases}
    \end{align*} 
    We use this path in the cases below to show there must be a non-causal path $u$ from $X_1$ to $Y$ in $\g[D]$ that is open given $\mb{Z_1}$, which by Theorem \ref{thm:as-graphical}, contradicts that $\mb{Z_1}$ is an adjustment set relative to $(X_1,Y)$.

    \textbf{CASE 1: Suppose there is no directed path} from $\{C_1, \dots, C_\ell\}$ to $X_1$ in $\g[D]$.
    Therefore, consider $C_1$ and its corresponding path $t_1$. Note that $q(X_1, C_1)$ and $t_1$ cannot share a node other than $C_1$, since any such node would have a directed path (possibly of length zero) in $\g[D]$ to $C_1$, $X_1$, or a collider on $q(X_1, C_1)$. And this would contradict the acyclicity of $\g[D]$, the fact that there is no directed path from $\{C_1, \dots, C_\ell\}$ to $X_1$ in $\g[D]$, or the definition of $C_1 \notin \An(\mb{Z_1}, \g[D])$ as the closest such collider to $X_1$ on $q$. Thus, we can consider the path $u := q(X_1, C_1) \oplus t_1$. Note that $q(X_1, C_1)$ is open given $\mb{Z_1}$ by the definition of $q$ and $C_1$. Further by the definition of $C_1$ and $t_1$, no node on $t_1$ is in $\mb{Z_1}$ and every node on $t_1$ is a non-collider on $u$. Therefore, $u$ -- a non-causal path in $\g[D]$ from $X_1$ to $Y$ -- is open given $\mb{Z_1}$, which is a contradiction.

    \textbf{CASE 2: Suppose there is a directed path} from $\{C_1, \dots, C_\ell\}$ to $X_1$ in $\g[D]$.
    Let $C_a$, $a \in \{1, \dots, \ell\}$, be the closest such collider to $Y$ on $q$, let $w$ be an arbitrary directed path from $C_a$ to $X_1$ in $\g[D]$, and let $A$ be the node on both $q$ and $w$ that is closest to $Y$ on $q$.

    When $A$ is a node on $q(C_\ell,Y)$, consider the path $u := (-w)(X_1, A) \oplus q(A, Y)$. Note by the definition of $C_a$ and $w$ that no node on $(-w)(X_1, A)$ is in $\mb{Z_1}$ and every node on $(-w)(X_1, A)$ is a non-collider on $u$. Further note that $q(A, Y)$ is open given $\mb{Z_1}$ by the definition of $q$ and $C_\ell$. Therefore, $u$ -- a non-causal path in $\g[D]$ from $X_1$ to $Y$ -- is open given $\mb{Z_1}$, which is a contradiction. 

    When $A$ precedes $C_\ell$ on $q$, let $C_b$ be the collider in $\{C_{a+1}, \dots, C_\ell\}$ closest to $A$ on $q(A,Y)$. Note by the definition of $C_a$ that $C_b \notin \An(X_1, \g[D])$. Therefore, we can consider the path $t_b$, which is directed from $C_b$ to $Y$. Note that $(-w)(X_1, A) \oplus q(A,C_b)$ and $t_b$ cannot share a node other than $C_b$, since any such node would have a directed path (possibly of length zero) in $\g[D]$ to $C_b$, $X_1$, or a collider on $q(A,C_b)$. And this would contradict the acyclicity of $\g[D]$, the definition of $C_a$ as the closest collider in $\{C_1, \dots, C_\ell\}$ to $Y$ on $q$ with a directed path to $X_1$, or the definition of $C_b \notin \An(\mb{Z_1}, \g[D])$ as the closest such collider to $A$ on $q(A,Y)$. Thus, we can consider the path $u := (-w)(X_1, A) \oplus q(A,C_b) \oplus t_b$. Note by the definition of $C_a$ and $w$ that no node on $(-w)(X_1, A)$ is in $\mb{Z_1}$ and every node on $(-w)(X_1, A)$ is a non-collider on $u$. The same holds for $t_b$ by analogous logic. Further, note that $q(A, C_b)$ is open given $\mb{Z_1}$ by the definition of $q$, $A$, and $C_b$. Therefore, $u$ -- a non-causal path in $\g[D]$ from $X_1$ to $Y$ -- is open given $\mb{Z_1}$, which is a contradiction.
\end{proofof}

% --- MINIMALITY CRITERIA ---%
\begin{proofof}[Lemma \ref{lem:minimal-criteria} (Probabilistic Criteria for Minimality)]
    Let $\mb{T}$ be an adjustment set relative to $(X, Y)$ in a causal model where $\mb{T} < X$. Then by Lemmas \ref{lem:helper1}-\ref{lem:helper2}, $\mb{T}$ is a minimal adjustment set relative to $(X, Y)$ if and only if \ref{lem:minimal-criteria-ai}-\ref{lem:minimal-criteria-aii} hold for all $T \in \mb{T}$.

    When \ref{lem:minimal-criteria-ai} or \ref{lem:minimal-criteria-aii} do not hold for some $T \in \mb{T}$, let $\mb{Z}$ be a proper subset of $\mb{T}$ such that \ref{lem:minimal-criteria-bi}-\ref{lem:minimal-criteria-biii} hold for all $Z \in \mb{Z}$. In addition to \ref{lem:minimal-criteria-biii}, note that trivially $X \ind \mb{Z} \,|\, \mb{T}$ and $Y \ind \mb{Z} \,|\, \mb{T} \cup \{X\}$, since $\mb{Z} \subset \mb{T}$. Therefore, by Theorem \ref{thm:c-equiv}, $\mb{Z}$ is c-equivalent to $\mb{T}$ and thus, is an adjustment set relative to $(X, Y)$. Then, as above, $\mb{Z}$ is minimal by \ref{lem:minimal-criteria-bi}-\ref{lem:minimal-criteria-bii} and Lemmas \ref{lem:helper1}-\ref{lem:helper2}.
\end{proofof}

% -------------  SUPPORTING RESULTS  -------------%
\subsection{Supporting Results}

% --- ELEMENTWISE MINIMAL ---%
\begin{definition}
\label{def:minimal-element}
{\normalfont (\textbf{Elementwise Minimal Adjustment Set})}
    A set $\mb{Z}$ is an \textit{elementwise minimal adjustment set} relative to $(\mb{X}, \mb{Y})$ if $\mb{Z}$ is an adjustment set relative to $(\mb{X}, \mb{Y})$ and if $\mb{Z} \setminus \{Z\}$ is not an adjustment set relative to $(\mb{X}, \mb{Y})$ for any $Z \in \mb{Z}$.
\end{definition}

% --- HELPER 1 ---%
\begin{lemma}[Criteria for Elementwise Minimality]
\label{lem:helper1}
    Let $\mb{Z}$ be an adjustment set relative to $(\mb{X}, \mb{Y})$ in a causal model where $\mb{Z} < \mb{X}$. Then $\mb{Z}$ is an elementwise minimal adjustment set relative to $(\mb{X}, \mb{Y})$ if and only if the following hold for some $X \in \mb{X}$, $Y \in \mb{Y}$, and for all $Z \in \mb{Z}$:
    \begin{enumerate}[label=(\roman*)]
        \item $X \notind Z \,\big|\, \mb{Z} \setminus \{Z\}$, \, and \label{lem:helper1-i}
        \item $Y \notind Z \,\big|\, \big[ \mb{Z} \setminus \{Z\} \big] \cup \{X\}$. \label{lem:helper1-ii}
    \end{enumerate}
\end{lemma}

\begin{proofofnoqed}[Lemma \ref{lem:helper1}]
    Let $\mb{X}$, $\mb{Y}$, and $\mb{Z}$ be pairwise disjoint node sets in a causal DAG $\g[D]$, where $\mb{Z}$ is an adjustment set relative to $(\mb{X}, \mb{Y})$ in $\g[D]$ and where $\mb{Z} < \mb{X}$.
    
    $\Rightarrow$
    Let $\mb{Z}$ be an elementwise minimal adjustment set relative to $(\mb{X}, \mb{Y})$, and consider an arbitrary $Z \in \mb{Z}$. Since $\mb{Z}$ is an adjustment set but $\mb{Z} \setminus \{Z\}$ is not an adjustment set relative to $(\mb{X}, \mb{Y})$, then by Theorem \ref{thm:as-graphical}, there must be a non-causal path $p$ from some $X \in \mb{X}$ to some $Y \in \mb{Y}$ in $\g[D]$ that contains $Z$ as a non-collider and is open given $\mb{Z} \setminus \{Z\}$. Note that \ref{lem:helper1-i} holds since $p(X,Z)$ is open given $\mb{Z} \setminus \{Z\}$. Further, \ref{lem:helper1-ii} holds since $p(Z,Y)$ is open given $\mb{Z} \setminus \{Z\}$ and does not contain $X$ so that $p(Z,Y)$ is open given $\big[ \mb{Z} \setminus \{Z\} \big] \cup \{X\}$.

    $\Leftarrow$
    Let \ref{lem:helper1-i} and \ref{lem:helper1-ii} hold for some $X \in \mb{X}$, $Y \in \mb{Y}$, and for all $Z \in \mb{Z}$. Then consider an arbitrary $Z \in \mb{Z}$. Note that by \ref{lem:helper1-i}, \ref{lem:helper1-ii}, and faithfulness, the following two paths must exist in $\g[D]$: \\[4pt]
        \centerline{ $p$: a path from $X$ to $Z$ that is open given $\mb{Z} \setminus \{Z\}$ and } \\ [4pt]
        \centerline{ $q$: a path from $Z$ to $Y$ that is open given $\big[ \mb{Z} \setminus \{Z\} \big] \cup \{X\}$ } 
        \centerline{ with the fewest colliders of all such paths.}

    Note that $p$ and $q$ share at least one node, since $Z$ is an endpoint on both paths. Therefore, let $M$ be the node on both $p$ and $q$ that is closest to $X$ on $p$, and define $s = p(X,M) \oplus q(M,Y)$. We show below that $s$ is non-causal, $q$ is open given $\mb{Z} \setminus \{Z\}$, and $M$ is a non-collider on $s$, where $M \notin \mb{Z} \setminus \{Z\}$. Since $p$ is open given $\mb{Z} \setminus \{Z\}$, it will immediately follow that $s$ -- a non-causal path from $X$ to $Y$ -- is open given $\mb{Z} \setminus \{Z\}$. Thus by Theorem \ref{thm:as-graphical}, $\mb{Z} \setminus \{Z\}$ is not an adjustment set relative to $(\mb{X}, \mb{Y})$. Since $Z$ was arbitrary, this will complete the proof.

    We start by showing that $s$ is non-causal. When $M=X$, then $X$ is a node on $q$. Since $q$ is open given $\big[ \mb{Z} \setminus \{Z\} \big] \cup \{X\}$, then $X$ must be a collider on $q$ so that $s=q(X,Y)$ is non-causal. When instead $M \neq X$, suppose for sake of contradiction that $p$ -- and therefore $s$ -- begins $X \to$. Since $\mb{Z} < \mb{X}$, then $p$ cannot be directed. Thus, $p$ contains a collider. But by the definition of $p$, the closest such collider to $X$ on $p$ must be in $\An(\mb{Z} \setminus \{Z\}, \g[D])$, which contradicts that $\mb{Z} < \mb{X}$. 

    Next, suppose for sake of contradiction that $q$ is blocked given $\mb{Z} \setminus \{Z\}$. Since $q$ is open given $\big[ \mb{Z} \setminus \{Z\} \big] \cup \{X\}$, there must be a collider on $q$ in $\An(X, \g[D]) \setminus \An(\mb{Z} \setminus \{Z\}, \g[D])$. Let $C$ be the closest such collider to $Y$ on $q$, and let $r$ be a directed path (possibly of length zero) from $C$ to $X$ in $\g[D]$. Note that $r$ and $q(C,Y)$ cannot share a node other than $C$, since any such node would have a directed path (possibly of length zero) to $C$, $Y$, or a collider on $q(C,Y)$. And this would contradict the acyclicity of $\g[D]$, the definition of $q$ as a path that is open given $\big[ \mb{Z} \setminus \{Z\} \big] \cup \{X\}$ with the fewest colliders, or the definition of $C \notin \An(\mb{Z} \setminus \{Z\}, \g[D])$ as the closest such collider to $Y$ on $q$.

    Thus, consider the path $t := (-r)(X, C) \oplus q(C,Y)$. Note by the definition of $q$, $C$, and $r$ that $t$ is a non-causal path from $X$ to $Y$ in $\g[D]$ that is open given $\mb{Z} \setminus \{Z\}$. By Theorem \ref{thm:as-graphical} and the fact that $\mb{Z}$ is an adjustment set relative to $(\mb{X}, \mb{Y})$, $t$ must be blocked given $\mb{Z}$, and therefore, $t$ must contain $Z$ as a non-collider. Note that $q(C,Y)$ cannot contain $Z$ by the definition of $q$ as a path from $Z$ to $Y$, and so $(-r)(X,C)$ must contain $Z$. But then $(-r)(Z, C) \oplus q(C,Y)$ is a path from $Z$ to $Y$ that is open given $\big[ \mb{Z} \setminus \{Z\} \big] \cup \{X\}$ with fewer colliders than $q$, which contradicts the definition of $q$. Therefore, $q$ is open given $\mb{Z} \setminus \{Z\}$.    
    
    Finally, we show that $M$ is a non-collider on $s$. It will follow that $M$ is a non-collider on either $p$ or $q$. Since $p$ and $q$ are open given $\mb{Z} \setminus \{Z\}$, then $M \notin \mb{Z} \setminus \{Z\}$. Thus, for sake of contradiction, suppose $M$ is not a non-collider on $s$. In each of the cases below, we find a non-causal path from $X$ to $Y$ that is open given $\mb{Z}$, which by Theorem \ref{thm:as-graphical} contradicts that $\mb{Z}$ is an adjustment set relative to $(\mb{X},\mb{Y})$.
    \begin{itemize}
        \item Let ${M=X}$. When ${M = X}$, we have already shown that $q(X,Y)$ is a non-causal path from $X$ to $Y$. Then note that $q(X,Y)$ is open given $\mb{Z}$, since $q$ is open given $\mb{Z} \setminus \{Z\}$ and $q(X,Y)$ does not contain $Z$.

        \item Let ${M=Y}$. When ${M \neq X}$, we have already shown that $p$ -- and therefore $p(X,Y)$ -- is a non-causal path from $X$ to $Y$. Then note that $p(X,Y)$ is open given $\mb{Z}$, since $p$ is open given $\mb{Z} \setminus \{Z\}$ and $p(X,Y)$ does not contain $Z$.

        \item Let $M$ be a collider on $s$. We have shown that $s$ is a non-causal path from $X$ to $Y$. To see that $s$ is open given $\mb{Z}$, note since $p$ and $q$ are open given $\mb{Z} \setminus \{Z\}$ that every collider on $s(X,M)$ and $s(M,Y)$ is in $\An\big( \mb{Z} \setminus \{Z\}, \g[D] \big)$ and no non-collider on $s(X,M)$ or $s(M,Y)$ is in $\mb{Z} \setminus \{Z\}$. Further, $s$ does not contain $Z$ except possibly $M=Z$. It remains to show that $M \in \An(\mb{Z}, \g[D])$. This clearly holds when ${M=Z}$ or when $M$ is a collider on $p$, since $p$ is open given $\mb{Z} \setminus \{Z\}$. Finally, consider when $M \neq Z$ and $M$ is a non-collider on $p$. Since $M$ is a collider on $s$, then $p(M,Z)$ begins $M \to$. Thus, either $p(M,Z)$ is directed or contains a collider in $\An(\mb{Z} \setminus \{Z\}, \g[D])$. \hfill \BlackBox
        \end{itemize}    
\end{proofofnoqed}

% --- HELPER 2 ---%
\begin{lemma}[Equivalency of Definitions \ref{def:minimal-adjust} and \ref{def:minimal-element}]
\label{lem:helper2}
    A set is an elementwise minimal adjustment set relative to $(X, Y)$ if and only if it is a minimal adjustment set relative to $(X, Y)$. 
\end{lemma}

\begin{proofof}[Lemma \ref{lem:helper2}]
    $\Leftarrow$
    Holds by definition.
    $\Rightarrow$
    Let $\{X,Y\}$ and $\mb{Z}$ be disjoint node sets in a causal DAG $\g[D]$, where $\mb{Z}$ is an elementwise minimal adjustment set relative to $(X,Y)$ in $\g[D]$. Let $|\mb{Z}| = n$, and note that the lemma holds trivially when $n \in \{0,1\}$. Thus, let $n \ge 2$. For sake of contradiction, suppose that $\mb{Z}$ is not a minimal adjustment set relative to $(X, Y)$, and let $\mb{Z_k}$ be a largest proper subset of $\mb{Z}$ such that $\mb{Z_k}$ is an adjustment set relative to $(X, Y)$. Let $|\mb{Z_k}| = k$, and note that $0 \le k \le n-2$. Then order the nodes in $\mb{Z} := \{Z_1, \dots, Z_n\}$ so that $\mb{Z} \setminus \mb{Z_k} = \{Z_{k+1}, \dots, Z_n\}$.

    Consider the set $\mb{Z_{k+1}} := \{Z_1, \dots, Z_{k+1}\}$. By the definition of $\mb{Z_k}$ as a largest subset, $\mb{Z_{k+1}}$ is not an adjustment set relative to $(X, Y)$. Thus by Theorem \ref{thm:as-graphical} and the fact that $\mb{Z}$ is an adjustment set, $\g[D]$ must contain a non-causal path $p$ from $X$ to $Y$ that is open given $\mb{Z_{k+1}}$. Note however that $p$ is blocked given $\mb{Z_k}$ by Theorem \ref{thm:as-graphical} and the definition of $\mb{Z_k}$ as an adjustment set. Therefore, $p$ must contain a collider in $\An(Z_{k+1}, \g[D]) \setminus \An(\mb{Z_k}, \g[D])$, where we use the convention that $\An(\emptyset, \g[D]) = \emptyset$. Of all such colliders, let $C_X$ be the closest to $X$ on $p$, and $C_Y$, the closest to $Y$ on $p$, where possibly ${C_X = C_Y}$. Define $r_x$ to be a directed path in $\g[D]$ (possibly of length zero) from $C_X$ to $Z_{k+1}$, and define $r_y$ analogously from $C_Y$ to $Z_{k+1}$. Then define $r_{xy}$ to be a longest directed path in $\g[D]$ (possibly of length zero) from $Z_{k+1}$ to $\{Z_{k+1}, \dots, Z_n\}$, and let $r_{xy}$ end with the node $Z_\ell$, $\ell \in \{k+1, \dots, n\}$.

    Since $\mb{Z}$ is elementwise minimal, then by Theorem \ref{thm:as-graphical}, we can define a non-causal path $q$ from $X$ to $Y$ that is open given $\mb{Z} \setminus \{Z_\ell \}$ and that contains $Z_\ell$ as a non-collider. Therefore, either $q(X, Z_\ell)$ ends $\gets Z_\ell$ or $q(Z_\ell, Y)$ begins $Z_\ell \to$. We show below that in both cases, there is a non-causal path $t$ from $X$ to $Y$ in $\g[D]$ that is open given $\mb{Z_k}$, which by Theorem \ref{thm:as-graphical} contradicts that $\mb{Z_k}$ is an adjustment set relative to $(X, Y)$ and completes the proof.\\

    \textbf{CASE 1: Suppose $q(X, Z_\ell)$ ends $\gets Z_\ell$.}
    To form $t$, we want to combine $q(X, Z_\ell)$, $(-r_{xy})(Z_\ell, Z_{k+1}) \oplus (-r_y)(Z_{k+1},C_Y)$, and $p(C_Y,Y)$.
    
    We start by supposing, for sake of contradiction, that $q(X, Z_\ell)$ contains a collider. Since $q(X, Z_\ell)$ ends $\gets Z_\ell$ and $q$ is open given $\mb{Z} \setminus \{Z_\ell \}$, it follows that $\g[D]$ contains a causal path $s$ from $Z_\ell$ to $\mb{Z} \setminus \{Z_\ell \}$. But then either $r_{xy} \oplus s$ contradicts the definition of $r_{xy}$ as a longest path from $Z_{k+1}$ to $\{Z_{k+1}, \dots, Z_n\}$, or $r_x \oplus r_{xy} \oplus s$ contradicts the definition of $C_X \notin \An(\mb{Z_k},\g[D])$.

    Thus, $q(X, Z_\ell)$ takes the form $X \gets \dots \gets Z_\ell$ so that $r := q(X, Z_\ell) \oplus (-r_{xy})(Z_\ell, Z_{k+1}) \oplus (-r_y)(Z_{k+1},C_Y)$ takes the form $X \gets \dots \gets C_Y$. If we let $B$ be the node on both $r$ and $p(C_Y, Y)$ that is closest to $Y$ on $p$, then we can define $t = r(X, B) \oplus p(B, Y)$.
    
     Note that $t$ is a non-causal path from $X$ to $Y$. To see that $t$ is open given $\mb{Z_k}$, note that no node on $t(X, B)$ is a collider on $t$. And no node on $t(X, B)$ is in $\mb{Z_k}$, since $C_Y$ is an ancestor of every node on $t(X, B)$ and $C_Y \notin \An(\mb{Z_k},\g[D])$. Finally, note by the definition of $p$, $C_Y$, and $B$ that every collider on $t(B, Y)$ is in $\An(\mb{Z_k},\g[D])$, and no non-collider on $t(B, Y)$ is in $\mb{Z_k}$.\\

    \textbf{CASE 2: Suppose $q(Z_\ell, Y)$ begins $Z_\ell \to$.}
    To form $t$, we want to combine $p(X, C_X)$, $r_x(C_X,Z_{k+1}) \oplus r_{xy}(Z_{k+1},Z_\ell)$, and $q(Z_\ell, Y)$.
    
    By analogous logic to CASE 1, $q(Z_\ell, Y)$ cannot contain colliders and thus, takes the form $Z_\ell \to \dots \to Y$. Then $r := r_x(C_X,Z_{k+1}) \oplus r_{xy}(Z_{k+1},Z_\ell) \oplus q(Z_\ell, Y)$ takes the form $C_X \to \dots \to Y$. If we let $B$ be the node on both $p(X, C_X)$ and $r$ that is closest to $X$ on $p$, then we can define $t = p(X, B) \oplus r(B, Y)$.

     For sake of contradiction, suppose that $t$ is causal, and consider where $B$ falls on $r$. If $B$ were to fall on $r(C_X,Z_\ell)$, then $t$ would contain $Z_\ell \in \mb{Z}$. But by Theorem \ref{thm:as-graphical}, this contradicts that $\mb{Z}$ is an adjustment set relative to $(X, Y)$. Thus, $B$ falls on $r(Z_\ell, Y)$. Note that $B \neq X$, since $r(Z_\ell, Y) = q(Z_\ell, Y)$, where $q$ is a path from $X$ to $Y$. This implies $B$ is a non-endpoint node on $p(X,C_X)$ and $p(X,C_X)$ begins $X \to$. But then $p(X,C_X)$ must contain a directed path from $B$ to either $C_X$ or a collider on $p(X,C_X)$. This path in combination with $r(C_X,B)$ contradicts either the acyclicity of $\g[D]$ or the definition of $C_X \notin \An(\mb{Z_k},\g[D])$ as the earliest such collider on $p$.

    To see that $t$ is open given $\mb{Z_k}$, note that no node on $t(B, Y)$ is a collider on $t$. And no node on $t(B, Y)$ is in $\mb{Z_k}$, since $C_X$ is an ancestor of every node on $t(B, Y)$ and $C_X \notin \An(\mb{Z_k},\g[D])$. Finally, note by the definition of $p$, $C_X$, and $B$ that every collider on $t(X, B)$ is in $\An(\mb{Z_k},\g[D])$, and no non-collider on $t(X, B)$ is in $\mb{Z_k}$.    
\end{proofof}

%%%%%%%%%%%%%%%%%%%%%%%%%%%%%%%%%%%%
%           SUPPLEMENT D             %
%%%%%%%%%%%%%%%%%%%%%%%%%%%%%%%%%%%%

\section{Additional Simulation Results}
\label{supp:simulations}

The following are two additional plots from our simulation results that we excluded from Section \ref{sec:simulations}. They consider Setting 3, where datasets are generated from models without a treatment effect. Compare these plots to Figures \ref{fig:sim-conclusions12}-\ref{fig:sim-effectdiffs12} that consider Settings 1-2, where datasets are generated from models with a treatment effect.

\begin{figure*}
    \centering
    \begin{tabular}{c}
        \includegraphics[width = .7\columnwidth]{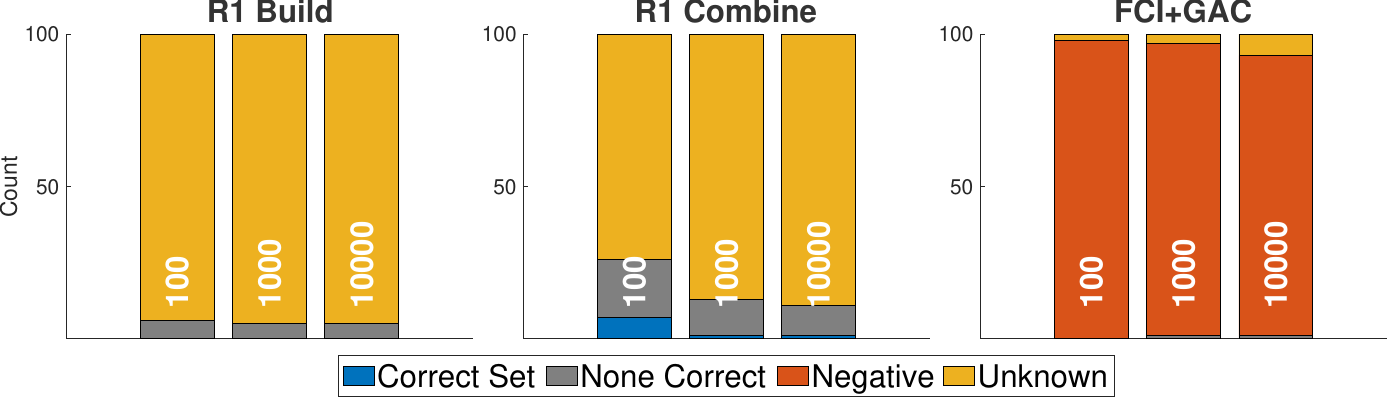}\\
        (a) Setting 3 \\ %{\color{lightgray} No Treatment Effect}\\
    \end{tabular}
    \caption{Outcomes of our data-driven methods (R1 Build, R1 Combine) and an existing graphical approach ({FCI + GAC}) on simulated data. Stacked bars show how often a method results in \textit{correct set}, \textit{none correct}, \textit{negative}, or \textit{unknown}.}
    \label{fig:sim-conclusions3}
\end{figure*}

Figure \ref{fig:sim-conclusions3} shows how often all three methods (R1 Build, R1 Combine, {FCI + GAC}) obtain one of the following outcomes.
\begin{itemize}[leftmargin=2.8cm,labelsep=0.3cm]
    \item[\textbf{Correct Set:}] Method concludes $\{X_1, X_2\}$ affects $Y$. At least one adjustment set it finds is correct \\ \hspace*{.5em} in the underlying DAG.
    \vskip .05in

    \item[\textbf{None correct:}] Method concludes $\{X_1, X_2\}$ affects $Y$. No adjustment set it finds is correct in the \\ \hspace*{.5em} underlying DAG.
    \vskip .05in
    
    \item[\textbf{Negative:}] Method concludes $\{X_1, X_2\}$ has no effect on $Y$.
    \vskip .05in

    \item[\textbf{Unknown:}] Method cannot find an adjustment set for the effect of  $\{X_1, X_2\}$ on $Y$.
\end{itemize}
Successful performance in Setting 3 is when a method obtains a \textit{negative} outcome. As noted in the main text, {FCI + GAC} easily outperforms R1 Build and R1 Combine in this setting, since our methods cannot obtain a \textit{negative} outcome.%
\footnote{For clarity in reading Figure \ref{fig:sim-conclusions3}, note that in a small percentage of datasets, R1 Combine incorrectly concludes that there is a non-zero treatment effect, but the adjustment set it finds is accurate (for a ``treatment effect'' of zero).}

\begin{figure*}
    \centering
    \begin{tabular}{c}
        \includegraphics[width = .7\columnwidth]{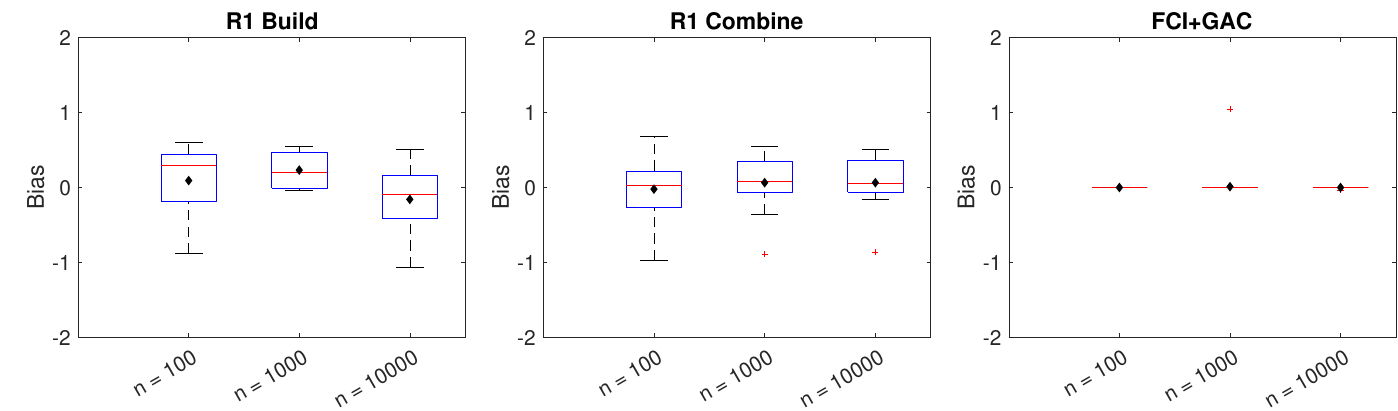}\\
        (a) Setting 3 \\ %{\color{lightgray} No Treatment Effect}\\
    \end{tabular}
    \caption{Differences between estimated and true causal effects. Results for our data-driven methods (R1 Build, R1 Combine) and an existing graphical approach ({FCI + GAC}) when applied to simulated data.}
\label{fig:sim-effectdiffs3}
\end{figure*}

%----------------

Figure \ref{fig:sim-effectdiffs3} shows the differences between causal effect estimates (based on conclusions from each method) and the true causal effect. As noted in the main text, {FCI + GAC} easily outperforms R1 Build and R1 Combine in this setting, since our methods cannot obtain a \textit{negative} outcome.

%----------------

\end{document}